# Generative AI, Managerial Expectations, and Economic Activity*

Manish Jha† — *Georgia State*
Jialin Qian‡ — *Rochester Inst. of Tech.*
Michael Weber§ — *Purdue and NBER*
Baozhong Yang¶ — *Georgia State*

November 2026

**Abstract**

Using generative AI, we extract managers' economic expectations from over 120,000 corporate conference call transcripts. The resulting AI Economy Score predicts GDP growth, production, employment, and wages five or more quarters ahead, beyond survey forecasts and other existing measures. Industry- and firm-level scores capture sector- and firm-specific activity, and a composite of firm, industry, and macroeconomic expectations further improves forecasting power and horizon. Predictability concentrates among historically more accurate, less biased managers, consistent with an information rather than a noise channel. Managerial expectations thus offer incremental insights into economic activity, with implications for macro- and microeconomic decision-making.

**JEL Classification**: C81, D2, E2, E3, E44

**Keywords**: Artificial Intelligence, ChatGPT, Generative AI, Large Language Model, Macroeconomic Forecast, Managerial Expectations, Textual analysis

---

*For helpful comments, we thank Vineet Bhagwat, Jiayin Hu, Haibo Jiang, Sophia Kazinnik, Sophia Zhengzi Li, Dexin Zhou, and participants at the NBER Forecasting & Empirical Methods, China International Conference in Finance, Conference on Financial Economics and Accounting, Indian School of Business Summer Research Conference, Financial Intermediation Research Society Conference, Northern Finance Association, Society of Quantitative Analysts Conference, Fidelity Quantitative Research, Georgia State University, International Conference on Information Technology and Quantitative Management, McMaster University, Peking University, Rutgers University, University of Oxford, and Winter Institute of AISSEAI. All errors are our own. An earlier version of this paper was circulated under the title "Harnessing Generative AI for Economic Insights."

†J. Mack Robinson College of Business, Georgia State University, Email: mjha@gsu.edu

‡Saunders College of Business, Rochester Institute of Technology, Email: jqian@saunders.rit.edu

§Purdue University Daniels School of Business, ESMT Berlin, and NBER, Email: weber366@purdue.edu

¶J. Mack Robinson College of Business, Georgia State University, Email: bzyang@gsu.edu

## 1. Introduction

Managers of publicly-listed corporations have a large impact on the economy because their corporations employ almost 30% of total non-farm employment in the U.S and are responsible for more than one quarter of GDP (Schlingemann and Stulz, 2022). For this reason, the government and public often listens carefully to their opinions and the media features them prominently.[1] It is well-documented that the expectations of households have important influences on future economic activities because consumption and other activities depend on their information and beliefs (e.g., Barsky and Sims, 2012; Chahrour and Jurado, 2018; Weber et al., 2022). By the same principle, managerial expectations should also contain valuable information for the economy at both the macro and micro level given their relevance for aggregate employment but also the fact that corporate investment is the most volatile component of aggregate GDP. However, surveys of managers are costly and challenging to conduct on a large scale, a constraint for their wide application.[2] In this paper, we apply the newly available generative AI tools to obtain managerial expectations on a wide range of economic factors from conference call transcripts. Relying on publicly available data, our approach provides an easily accessible set of expectation variables to researchers and policy makers.

We construct managerial expectations from conference call transcripts, using GenAI tools (specifically, GPT-5 mini). In particular, we feed corporate conference call transcripts to a generative AI model and prompt it to provide answers regarding managerial expectations of future states of the economy and the firm. We do this in two passes. The first pass reads the transcript in segments and retains only the statements in which management speaks to conditions of the US economy. The second pass reads the retained statements for the whole call and scores, on a seven-point scale, how optimistic management is about US economic conditions in the coming quarter. Separating extraction from scoring matters because an earnings call is long and mostly about the firm, so a model asked to do both at once has to find the macroeconomic content and judge it simultaneously, with the relevant material diluted by a large volume of firm-specific text. This approach generates a panel of answers regarding Gross Domestic Product (GDP) growth, production, employment, and investment, from more than 120,000 conference calls, representing 5,513 unique companies. We then aggregate these answers at the national and industry-sector levels to obtain economic measures that can be indicative of economic activities at these levels.

[1]For example, the Wall Street Journal highlighted JPMorgan Chase Chief Executive Jamie Dimon's perspective on U.S. interest rates (Saeedy, 2024). Additionally, CEOs of major companies such as Blackstone, Dell, Disney, Pepsi, and Tesla frequently serve on the President's economic advisory council (Gernon, 2017).

[2]Some examples include the CFO survey, conducted by Duke University's Fuqua School of Business and the Federal Reserve Banks for Richmond and Atlanta: https://cfosurvey.fuqua.duke.edu/; and the Survey of Firms' Inflation Expectations (SoFIE), conducted by Federal Researve Bank of Cleveland: https://www.clevelandfed.org/indicators-and-data/survey-of-firms-inflation-expectations. Important contributions on how firms form their expectations are Graham and Harvey (2001); Ben-David, Graham, and Harvey (2013); Coibion, Gorodnichenko, and Kumar (2018); Coibion, Gorodnichenko, and Ropele (2020). Candia, Coibion, and Gorodnichenko (2023) provide an excellent recent overview of this literature.

We first focus on the main measure, the *AI Economy Score*, which captures the average managerial expectation for the US economy in the next quarter. The *AI Economy Score* is strongly predictive of next quarter's real GDP, increasing the share of the variation explained by close to five percentage points in a model with common predictors for aggregate growth including the *Term Spread*, the real federal funds rate (*Real FFR*), *GZ Spread* (Gilchrist and Zakrajšek, 2012), and lag of GDP. This predictive power is also moderately persistent, lasting for five quarters. The *AI Economy Score* is also significantly associated with other future aggregate economic variables, including industrial production, employment, and wages, for 5 to 10 quarters in the future.

To filter out the influence of other current economic factors, we employ a vector autoregression (VAR) framework to study the impulse responses to changes in the managerial expectation score. We find that shocks to the *AI Economy Score* orthogonal to current economic activities significantly influence future activities. A positive shock to the expectation score predicts higher consumption, investment, and output growth for 8 or more quarters. It is also positively associated with higher inflation, excess market returns, and long-term government bond yields. The increased economic activity induced by the shock also triggers tighter monetary policy.

Our approach also allows the construction of more micro-level expectation variables, such as an AI-based industry-level economy score, by aggregating firm-level measures to the sector level. In the empirical work we measure the industry and firm scores net of the national average and of the sector average, respectively, so that the three levels are orthogonal and their coefficients are separately identified. Such scores exhibit substantial cross-sectional heterogeneity, with both pessimistic and optimistic outlooks for different industries at the same point in time. The aggregate and industry economy scores possess independent predictive power of future economic activities across industry sectors. In fact, the predictive power of the industry-level scores for future GDP growth lasts 4 years, suggesting that industry-level expectations have the potential to aid long-term economic decisions.

Furtermore, we can go one step further and zoom in at the most micro-level, i.e., consider the firm-level managerial expectations in predicting firm-level outcomes, including value-added, sales, employment, and wages. Both AI-based industry-level and firm-level scores forecast future firm-level activities, with firm-level expectations carrying predictive power for up to 4 years.

Our methods also enable the direct construction of managerial expectations of a host of economically related variables, such as employment, wages, industrial production, and others. We verify these expectation scores provide significant forecasting power for the corresponding future economic variables at both the aggregate and industry levels, and in both the short and long term. While these expectation scores are correlated with the AI economy scores, these variables contain additional information. For example, AI employment scores are more powerful predictors of future employment and unemployment rates than the overall economy score.

Morever, beyond explicit forecasts, managerial expectations about other economic variables also contain implicit signals about future economic growth. To extract this information, we prompt ChatGPT with 14 different questions covering various aspects of managers' expectations, from firm financial prospects to macroeconomic trends. We aggregate firm-level AI scores derived from ChatGPT responses into industry and national-level scores. A new measure, the *AI Weighted Score*, is constructed using the weights from a firm-quarter-level regression model that relates these scores to future firm sales. Results show that this score meaningfully extends the horizon of GDP predictability at the national level: it remains a significant predictor of real GDP growth through eight quarters ahead with higher predictive power than the original AI Economy Score in the long horizon. These findings suggest that leveraging AI to extract multidimensional managerial expectations from earnings calls can enhance economic forecasting, particularly for long-term growth projections.

Why are managerial expectations so powerful in forecasting future economic activities? One important reason is that managers can act upon their beliefs and their actions carry significant economic impacts. For example, Jha, Qian, Weber, and Yang (2023) show that managerial expectations of corporate investment policies significantly affect long-term investment activities. Another potential reason is that managers are privy to first-hand, detailed economic data and information that may not be available to the public. Therefore, aggregating such information and insights from managers would be particularly fruitful. Overall, our study shows that the new AI tools allow easy extraction of managerial expectations on a whole range of economic indicators, providing a new set of forward-looking information to researchers, investors, and regulators. While such information is mostly positive in nature, given its significant predictive power, it can also be instrumental for normative policies.

To better understand the source of this predictive power, we dissect the managerial forecasts to account for behavioral biases and firm-level heterogeneity that may influence forecast quality. We first investigate how innate managerial disposition affects foresight by classifying executives as Optimists, Pessimists, or Realists based on their historical forecasting patterns relative to realized economic outcomes. We also consider the historical accuracy of managerial expectations. This analysis reveals that the predictive power of managerial forecasts is concentrated among managers who are historically accurate or unbiased (Realists), whereas forecasts from historically inaccurate managers contain little incremental predictive content. The evidence is consistent with the information interpretation and speaks to the *news* vs *noise* debate in the macroeconomics literature on managerial expectations (e.g., Barsky and Sims, 2011; Chahrour and Jurado, 2018; Schmitt-Grohé and Uribe, 2012; Blanchard, L'Huillier, and Lorenzoni, 2013), thanks to the availability of micro-level managerial expectations data we constructed.

We then examine how a firm's operational structure shapes its manager's perspective, testing whether diversification provides a clearer or more muddled economic view. Our findings indicate that forecasts from managers at more focused firms, or those with lower business diversification, are significantly more powerful, suggesting that deep, specialized knowledge provides a stronger economic signal.

Finally, we conduct a host of robustness tests as well as address the potential look-ahead bias in applying large-language models. First, in a "horse race" against the Survey of Professional Forecasters (SPF), the *AI Economy Score* retains its predictive power and is comparable in strength to the survey-based forecast, with both measures remaining significant when included in the same model. Second, to address potential "look-ahead bias," we anonymize transcripts, i.e., removing all firm and date information, following Engelberg, Manela, Mullins, and Vulicevic (2025), and find the resulting score based on the anonymized texts remains a significant predictor. Third, we conduct an out-of-sample test using only conference calls from 2024Q2 to 2025Q4, a period that falls after the GPT-5 mini training data cutoff, and confirm that the firm-level score still robustly predicts future firm sales and the industry-level score predicts industry real GDP. Finally, we replicate the analysis using Google's open-source Gemma model, showing that our results are not dependent on a specific generative AI model.

Our paper contributes to the extensive literature on economic expectations and real economic activities. Related studies have examined how news about future economic conditions impacts business cycles and macroeconomic stability (Barsky and Sims, 2011; Chahrour and Jurado, 2018; Schmitt-Grohé and Uribe, 2012; Blanchard, L'Huillier, and Lorenzoni, 2013). Other research has explored how changes in consumer confidence, firm expectations, and perceptions of monetary policy influence economic activities (Barsky and Sims, 2012; Bauer, Pflueger, and Sunderam, 2024; Coibion, Gorodnichenko, and Kumar, 2018; Coibion, Gorodnichenko, and Weber, 2022), that financial-market risk perceptions predict macroeconomic outcomes (Pflueger, Siriwardane, and Sunderam, 2020), and how survey data on subjective beliefs affect the business cycle and determinants of inflation expectations (Bhandari, Borovička, and Ho, 2022; Coibion, Gorodnichenko, and Ropele, 2020; Weber, D'Acunto, Gorodnichenko, and Coibion, 2022). Our managerial expectation measures are strongly predictive of future economic activities. They can serve as complements to the standard consumer surveys and surveys of professional forecasters.

A related literature provides variables helping to predict economic activities , e.g., Gilchrist and Zakrajšek (2012). A closely related paper (Bybee, 2025) generates survey results from Wall Street Journal news articles. We regard the two papers as complementary since we focus on the expectations retrieved from corporate managers whereas the focus in Bybee (2025) is on news reports. Furthermore, Bybee (2025) allows LLMs to make inferences about macroeconomic states based on news articles, which involves more reasoning of the model, whereas in our paper the role of generative AI is more passive – to extract the managerial expectations expressed in the conference calls. Our approach generates variables with long-run predictability and also has the capability of generating aggregate, industry-level, state-level, and firm-level expectations, which can be fruitfully employed both at the macro- but also at the micro-level. Additionally, our use of conference call transcripts to gauge national sentiments sets this work apart from other works that rely on book corpora (Jha, Liu, and Manela, 2022) or newspapers (van Binsbergen et al., 2024).

Finally, our paper expands the use of generative AI tools in economics.[3] Korinek (2023) provides a survey of potential uses of generative AI in economic research. Our study and method may help to open more avenues for economic research with the help of AI. For example, a recent publication by Sheng, Sun, Yang, and Zhang (2024) has built on our approach to investigate the reliance on generative AI tools by investment companies in their investment strategies.

[3]This is a fast-growing literature. A partial list includes: Lopez-Lira and Tang (2023); Hansen and Kazinnik (2023); Kim, Muhn, and Nikolaev (2023, 2024); Jha, Qian, Weber, and Yang (2023); Yang (2023); Li, Mai, Shen, Yang, and Zhang (2023); Ouyang, Yun, and Zheng (2024).

## 2. Data

In this section, we discuss the different datasets we use as well as the variable construction.

### 2.1. Conference Call Transcripts

We obtain earnings call transcripts for publicly listed firms spanning the years 2006 to 2020 from the SeekingAlpha website and for the period from 2021 to 2025 from the FinancialModelingPrep website. These transcripts are derived from quarterly earnings call conferences, which are critical events in which firms engage with their investors and analysts following the announcement of their quarterly financial performance. During these calls, managers typically begin with a detailed speech that summarizes the firm's recent performance, discusses their outlook, and addresses any potential concerns regarding the firm's future. After the initial presentation, the floor is opened to analysts and investors, who then ask questions. These questions often pertain not only to the specifics of the firm's operations but also touch on broader economic conditions. Managers respond to these questions in real time, providing their insights and opinions. This interaction offers a valuable window into the perspectives of top management on both their company's prospects and the general economic environment.

### 2.2. Real Outcomes

We collect national-level data on GDP, industrial production, employment, federal funds rates, and treasury yields from the Federal Reserve Economic Data (FRED) database managed by the Federal Reserve Bank of St. Louis. We obtain industry-level outcomes from the U.S. Bureau of Economic Analysis (BEA) GDP by Industry accounts. Sectoral *Real GDP* is real value added by industry, measured in chained dollars, seasonally adjusted, and available quarterly at the NAICS sector level. Industry employment and wages are the BEA's annual series on full- and part-time employees and on wages and salaries by industry, which are not available at a quarterly frequency. Additionally, we gather firm-level financial outcomes from Compustat and the Center for Research in Security Prices (CRSP).

### 2.3. Survey Data

We use forecasts on economic indicators from the Survey of Professional Forecasters website (SPF) conducted by the Federal Reserve Bank of Philadelphia. Established in 1968, the SPF is one of the oldest quarterly surveys of its kind in the United States. The collected data provides valuable insights into the collective expectations of professional forecasters, helping policymakers, researchers, and the public understand economic trends and make informed decisions. In Section 7.1, we conduct a horserace analysis, comparing the professionals' forecasts with those made by firm managers as retrieved by generative AI.

### 2.4. Economic Target Variables

Our final sample comprises 125,830 earnings call transcripts spanning the years 2006 to 2025. We aggregate managerial forecasts to the national and industrial level and examine the predictive ability of these measures on future economic development across different horizons. In all levels of analyses, we focus on forecasting the following common economic indicators: GDP, production, employment, and wages. For firm-level predictions, we use *Value-Add*, or the difference between sales and cost of goods sold, as a proxy for the overall value created by firms and *Sales* as a proxy for production.

We define the realized economic indicators as the logarithm of economic indicators for the next period $i$ we forecast $(t+i)$ over the most recent period with known predictors $(t-1)$ in our regression analyses. We control for common economic predictors in line with Gilchrist and Zakrajšek (2012): *Term Spread*, *Real FFR* and *GZ Spread*. *Term Spread* is the difference between ten-year and three-month constant-maturity Treasury yield. *Real FFR* is the real federal funds rate, earlier work including Bernanke and Blinder (1992) show that the federal funds rate offers insights into upcoming shifts in real macroeconomic factors. Additionally, the *GZ Spread* serves as the indicator of movements in credit spreads.

## 3. Empirical Methodology

In this section, we describe the construction of our key AI-based economic forecast measure, provide validations of the measure, and present summary statistics.

### 3.1. AI Economy Score

We use OpenAI's GPT-5 mini model to process texts.[4] We construct the score in two passes. An earnings call runs to roughly 7,500 words, and most of what managers say concerns the firm rather than the economy. Asking a model to read an entire call and return a single number requires it to do two things at once, locate the macroeconomic content and judge it, with the relevant material diluted in a large volume of firm-specific text. We separate the two tasks. The first pass extracts the statements in which management speaks about US economic conditions. The second pass reads only those statements and produces the score.

In the first pass, we split each transcript into segments of approximately 16,000 characters, breaking at sentence boundaries, and issue one extraction request per segment. The model returns, verbatim, the sentences in which management expresses a view or expectation about the US economy rather than about the firm alone. Qualifying content includes, for example, commentary on US macroeconomic conditions, consumer demand and confidence, employment, interest rates, inflation, recession risk, tariffs, and housing. Backward-looking company results, firm-specific operational updates, conversational filler, operator lines, and disclaimer boilerplate are excluded, and the model is instructed to return nothing rather than stretch a marginal sentence. We then apply a mechanical filter that removes the negative responses the model returns for segments with no qualifying content, concatenates the surviving sentences across segments in their original order, and drops calls for which no content remains.

In the second pass, we assemble the surviving sentences for a call into a single numbered list and issue one request per conference call. Because the extraction output is compact, the scoring model sees the entire call's macroeconomic content at once. Each call generates exactly one request, and the score is defined at the firm-quarter level by construction, with no aggregation across pieces of a transcript. We supply the calendar quarter and the call date, and ask how optimistic or pessimistic management is about US economic conditions in the quarter immediately following the call. The model returns an integer together with a one-sentence explanation. The Internet Appendix reports both prompts in full.

The *AI Economy Score* for a firm-quarter is the integer the model returns, which runs from 1 for a severe expected deterioration in US conditions to 7 for a strong expected expansion, with 4 as the neutral midpoint. Three features of the design are worth noting. First, the prompt permits the model to infer management's view from indirect commentary rather than requiring an explicit statement about the economy. Managers rarely deliver macroeconomic forecasts as such. They talk about what their customers are doing, what they are paying for inputs, and whether they can hire, and a rule that required the word "economy" would discard most of the available signal. Second, the neutral value is reserved for cases in which the positive and negative evidence are genuinely offsetting, and the prompt asks for the adjacent values only when the balance clearly leans one way, which is what gives the aggregated series its time-series variation. Third, the prompt supplies the call date and instructs the model to reason using only what it would have known at that date. This feature bounds the model's use of hindsight but does not eliminate the concern, since the model is told when the call took place and its training data extend past that point. We therefore do not rely on the instruction alone. To further address potential look-ahead bias in applying large-language models, Section 7.2 re-estimates the measure on transcripts from which all firm and date information has been removed, and Section 5 re-estimates it on calls that postdate the model's training window entirely.

[4] In earlier versions of this study, we used GPT-3.5 Turbo and obtained comparable results. We adopt GPT-5 mini as our main model because it is more recent and more capable, and the consistency across model generations suggests our measure is robust to the choice of the underlying language model.

When a call contains nothing US-economy related to reason from, the model returns a null score and we drop that call from the quarter. We then aggregate the firm-quarter scores to the national level by taking the equally weighted average across all firms with a valid score in a quarter, and to the industry level by taking the equally weighted average across firms within a NAICS sector.

### 3.2. Validation

Validating machine learning outputs is crucial for ensuring accuracy and reliability. We employ several methods to validate the responses generated by ChatGPT: (i) Manual review and frequency analysis: We manually read through a random sample of around one thousand conference calls and ChatGPT responses to verify the accuracy and relevance of the generated content. We also examine the most commonly used words associated with both low and high AI Economy Scores to gauge the consistency and appropriateness of ChatGPT's outputs; (ii) Trend analysis: We analyze trends over time and across various industries to assess the coherence and relevance of ChatGPT's responses within different contexts.

**3.2.1. Manual Review and Frequency Analysis.** We first manually go over a random sample of the conference calls and ChatGPT responses, verifying the accuracy and relevance of the generated content. We find ChatGPT to consistently generate high-quality responses similar to human assessments. Internet Appendix provides example texts associated with high and low AI scores respectively.

Next, we analyze the textual explanations behind the scores. The second pass of our process returns a one-sentence explanation alongside each score. We analyze these explanations in two ways. First, we collect the explanations attached to positive changes in managerial optimism, and separately those attached to negative changes, group each set into topics, and plot the resulting shares by year in Figure 1. Consumer demand is the largest category on both sides, at 27.0% of the positive explanations and 32.3% of the negative ones, followed by macroeconomic conditions among the positive and by policy, regulatory, and geopolitical uncertainty among the negative. The composition is otherwise stable across the sample, with one exception: public health crises, negligible in every other year, account for a large share of the negative explanations in 2020 and 2021, consistent with the impact of the COVID-19 pandemic.

[Insert Figure 1 Here]

Second, we extract frequently appearing 3-grams and 4-grams from the explantion texts for low- and high-score transcripts to understand the economic contexts. We split the explanations into a low bucket, containing the calls scored 1 to 3, and a high bucket, containing the calls scored 5 to 7. The neutral score of 4 is excluded.[5] Rather than ranking phrases by raw frequency, which would surface language common to both buckets, we rank them by distinctiveness, defined as the difference between a phrase's frequency per 1,000 explanations in its own bucket and its frequency per 1,000 explanations in the opposite bucket. A phrase must appear at least 20 times in its own bucket to be eligible.

[Insert Table 1 Here]

[5] From the explanation texts, we remove stopwords and non-English words using the dictionaries available in the Python Natural Language Toolkit package. We remove proper nouns and then lemmatize each word to account for differing grammatical noun and verb forms. We also drop two classes of phrases that appear frequently in both buckets and therefore carry no information about direction: the grading vocabulary the model uses to justify a score, such as "optimistic" or "management expects," and neutral magnitude phrasing such as "single digit." We then use the CountVectorizer function in the Python package sklearn to count 3-grams and 4-grams in each bucket.

Table 1 shows the top 10 phrases (3- and 4-grams) associated with low and high AI Economy scores. The phrases in both columns are macroeconomic rather than firm-specific, which is what the prompt asks the model to attend to. For low AI Economy scores, the leading phrases point to weak or softening final demand ("weak consumer demand," "weak consumer spending," "soft consumer demand," "consumer discretionary spending"), to supply-side disruption and cost pressure ("supply chain disruption," "ongoing supply chain disruption," "persistent inflation supply chain," "inflation drive cost pressure"), to the interaction of inflation with monetary policy ("inflation rise interest"), and to policy uncertainty ("tariff driven uncertainty," "tariff drive cost pressure"). Longer sales cycles also appear, in "elongate sale cycle," which is how a slowdown in business spending shows up in managers' language before it shows up in the aggregate data. In contrast, the phrases associated with high AI Economy scores describe strong and broad-based demand ("strong consumer demand," "robust consumer demand," "strong broad base demand," "continue strong consumer demand"), regional strength ("strong north american demand"), and tight labor markets ("tight labor market"). Notably, "supply chain" and "inflation" phrases appear in both columns but with different valence: in the low bucket they describe disruption and cost pressure, whereas in the high bucket they appear alongside demand strength, as in "labor shortage supply chain" and "input cost inflation," which is the language of an economy running hot. The contrast is consistent with the model separating the direction of macroeconomic change from the topic being discussed.

**3.2.2. Trend Analysis.** We next compare the time trend in the *AI Economy Score* with realized GDP growth and with the GDP forecast from the Survey of Professional Forecasters (SPF). Panel (a) of Figure 2 plots the score against the change in realized GDP, measured as the percent change in real GDP in quarter $t+1$ relative to quarter $t-3$. The two series track each other closely across the sample. The score turns down ahead of the 2008 financial crisis, well before the trough in realized GDP. It does not anticipate the collapse in output in the first quarter of 2020, falling alongside realized GDP rather than ahead of it, which is unsurprising given how little warning the pandemic gave. It does capture the recovery, spiking with realized GDP growth in 2021 before both series settle back to pre-pandemic levels over 2022–2025. Panel (b) repeats the comparison against the SPF forecast, where the two series move together just as closely: the score again turns down earlier in 2007–08, and both trace the same collapse and rebound in 2020–21.

Panel (b) illustrates the time-series trends of the *AI Economy Score* alongside the change in the SPF forecast, measured as the SPF-based GDP forecast for quarter $t+1$ relative to realized GDP in quarter $t-3$, over the entire sample period. These two lines also demonstrate strong comovement, including during the COVID-19 downturn and the 2021 rebound. Since the AI economy score aggregates opinions of corporate managers, its high correlation with the SPF forecasts suggests that managers and professional forecasters indeed have similar outlooks on the economy. However, we note that the AI forecasts possibly carry additional information beyond those in the SPF forecasts given the differences in the information sets and skillsets of managers and forecasters (Section 7.1).

[Insert Figure 2 Here]

Next, we analyze the heterogeneity in the *AI Economy Score* across industries. Figure 3 plots the average yearly *AI Economy Score* from 2006 to 2025 for the 18 North American Industry Classification System (NAICS) sectors represented in our sample; the nineteenth, Management of Companies and Enterprises, contains no firm in any year. During the recession of 2007–2009, below-average *AI Economy Scores* arise across virtually all industries, highlighting the breadth of that downturn. At the annual frequency, the 2020 recession appears more muted: below-average scores are visible in only a few sectors, such as Mining, Oil and Gas, reflecting the collapse in energy prices, while the sharp intra-year swings in 2020 largely offset one another in the yearly averages.

[Insert Figure 3 Here]

Substantial heterogeneity for AI forecasts across industries is present. During the 2007–2009 recession, the Retail Trade sector was most severely affected, alongside Accommodation and Food Services, Arts and Recreation, Finance and Insurance, and Real Estate. Weakness in the Educational Services and Health Care sectors instead emerged with a lag, with depressed scores persisting into 2011–2013. In contrast, the Construction and Agriculture sectors were relatively resilient during the crisis itself. Post-recession recovery is evident in the above-average *AI Economy Scores*, shown in blue. In the years following the 2007–2009 recession, the Construction sector exhibited the highest and most persistent levels of optimism, with scores well above the sample average from 2012 through 2019, followed by Transportation and Warehousing. After the COVID-19 pandemic, 2021 stands out as a broad-based column of optimism across nearly all industries, most pronounced for the sectors hardest hit by pandemic restrictions, including

Arts and Recreation, Accommodation and Food Services, and Construction. Pockets of renewed pessimism appear in 2022–2023, notably in Retail Trade. These findings align well with prior expectations and actual historical events.

### 3.3. Summary Statistics

Table 2 displays the descriptive statistics of the *AI Economy Score*, the realized values of various macroeconomic indicators, and predictor variables of future economic conditions at the aggregate level.

[Insert Table 2 Here]

Over the 80 quarters in our sample period (2006–2025), the average *AI Economy Score* is 4.252, with a standard deviation of 0.427, a median of 4.331, and an interquartile range from 4.118 to 4.479. The mean sits just above the neutral midpoint of 4, so the average manager in the average quarter expects US conditions to improve slightly over the coming quarter. The median exceeds the mean, which reflects the fact that the series is pulled down by a small number of severe quarters rather than by a general pessimism. The dispersion is economically meaningful for a scale of integers: a one-standard-deviation move in the national average corresponds to roughly four in ten firms shifting by a full point on the seven-point scale in the same direction. The entity-neutered version of the score, constructed on anonymized transcripts in Section 7.2, has a lower mean of 3.994 and a wider standard deviation of 0.620, which is consistent with the model having less context to work with once identifying information is stripped out. In our predictive regressions, we use the logarithm growth rates of the economic indicators (including *Real GDP*, *Industrial Production*, *Payroll Employment*, and *Wages*) for the next period over the most recently known period. *Term Spread* and *Real FFR* are quarterly measures obtained from averaging monthly values. The credit spread index *GZ Spread* is the quarterly GZ spread obtained by taking the average of the monthly values.

## 4. AI Forecasts on Economic Indicators

### 4.1. AI-Predicted Economy Score and Realized Real GDP

To evaluate the predictive ability of the AI-Predicted Economy Score for economic activities, we estimate the following forecasting specification:

$$ln\frac{Real\ GDP_{t+1}}{Real\ GDP_{t-1}} = \alpha + \beta_1 AI\ Economy\ Score_t + \beta_2 Term\ Spread_t + \beta_3 Real\ FFR_t + \beta_4 GZ\ Spread_t + \sum_{i=1}^{4} ln\frac{Real\ GDP_{t-i}}{Real\ GDP_{t-i-1}} + \varepsilon_t \quad (1)$$

where *Term Spread* is the difference between ten-year and three-month constant-maturity Treasury yield, *Real FFR* denotes real fed funds rate, and *GZ Spread* is the credit spread index. Four lags of *Real GDP* growth are included as controls.

Table 3 details the comparison of the predictive power of two financial indicators, the *GZ Spread* and the managerial economic forecast, *AI Economy Score*, for real GDP growth. We focus on the one-quarter and longer-term (up to 6 quarters ahead) forecast horizons and report the coefficients associated with the financial indicators, as well as the goodness of fit as measured by $R^2$.

Panel A in Table 3 presents the results for the one-quarter ahead forecasting period. Column (1) shows that the *GZ Spread* has strong explanatory power for real GDP, consistent with Gilchrist and Zakrajšek (2012), whereas the *Term Spread* and *Real FFR* do not exhibit short-run forecasting ability during our sample period. In column (2), we replace the *GZ Spread* with the *AI Economy Score*. The coefficient on the *AI Economy Score* is positive and significant at the 1% level, indicating that our *AI Economy Score* is a strong predictor of the next quarter's real GDP growth. The effect is also economically significant - one standard deviation increase in the *AI Economy Score* implies a 1.30 percentage point (0.0304 × 0.427) increase in the growth rate of real GDP over the subsequent quarter. In column (3), we include both the *GZ Spread* and the *AI Economy Score* in the predictor set. The *AI Economy Score* remains statistically significant at the 1% level. The $R^2$ increases from 0.438 in column (1) to 0.486 in column (3), indicating that the *AI Economy Score* adds substantial incremental forecasting power to those of existing predictors.

In Panel B, we examine the long-term predictive power of the *AI Economy Score*. Our analysis reveals that the *AI Economy Score* retains its predictive power for real GDP growth for up to five quarters. In contrast, the *GZ Spread* remains insignificant through the fourth quarter and becomes weakly significant in the fifth and sixth quarters. Our results indicate that the *AI Economy Score* provides valuable information about long-term economic prospects that extends beyond the scope of existing predictors.

These results show that the *AI Economy Score* contains additional information about the future economy and provides extra predictive power for real GDP, beyond the common predictors documented in the literature (Bernanke and Blinder, 1992; Gilchrist and Zakrajšek, 2012; Saunders et al., 2025).

### 4.2. AI-Predicted Economy Score and Various Economic Indicators

As managers usually touch on multiple topics related to the economy during earnings conference call, the economic forecasting score extracted from these calls can contain forecasts on different aspects of economic activities. In Table 4, we test the forecasting ability of the *AI Economy Score* on the growth of Industrial Production, Employment, and Wages over the short and long term with the same specifications as in Table 3. In columns (1), (4), and (7) of Panel A, we include the three existing economic predictors: *Term Spread*, *Real FFR*, and *GZ Spread*. The results show that the *GZ Spread* is a strong predictor of Industrial Production for the next quarter, whereas the *Term Spread* and *Real FFR* predict only Industrial Production, and not Employment or Wages. We replace the *GZ Spread* with *AI Economy Score* in columns (2), (5), and (8) to compare the predictive power of the two variables. The goodness-of-fit comparison implies that the *AI Economy Score* exhibits higher predictive power for Employment and Wages, whereas the *GZ Spread* performs better for Industrial Production. For example, the inclusion of the *AI Economy Score* to predict employment yields an $R^2$ of 31.1%, a 18.7% improvement compared to that of the *GZ Spread*. Also, simultaneously adding both the *GZ Spread* and *AI Economy Score* in the model reduces the predictive power of the *GZ Spread*, as shown in columns (3), (6), and (9). The magnitude of the coefficients attached to the *AI Economy Score* is also economically significant: a one-standard-deviation increase of *AI Economy Score* is associated with 2.67%, 1.23%, and 0.21% increases in the growth rate of industrial Production, Employment, and Wages over the next quarter, which corresponds to 83%, 59%, and 42% of a standard deviation in the growth rate of Industrial Production, Employment, and Wages.

[Insert Table 4 Here]

In Panel B, we study the longer term predictive power of the *AI Economy Score* on industrial production, employment, and wages. The results imply that the *AI Economy Score* predicts industrial production for about five quarters, although the statistical power weakens for predictions four to five quarters ahead. Additionally, the results show that the *AI Economy Score* predicts employment for up to five quarters and wages for approximately ten quarters. In contrast, while the *GZ Spread* predicts industrial production for four quarters and employment for eight to ten quarters ahead, it fails to predict employment at horizons shorter than eight quarters and shows only sporadic, weak significance for wages.

Table 4 therefore shows the *AI Economy Score* predicting industrial production, employment, and wages over long horizons, controlling for known predictors, and doing so in a way that complements rather than duplicates the *GZ Spread* for employment and wages. This is plausible because the AI-predicted score is obtained from managers' outlooks and conversations with investors and analysts, and it should contain more information about firms' labor-related policies as compared to conditions in credit markets.

### 4.3. VAR Analysis

To separate the information in managerial expectations from current economic and financial conditions, we follow the VAR of Gilchrist and Zakrajšek (2012) and replace their excess bond premium with the *AI Economy Score*. The endogenous variables, in order, are: (a) real personal consumption expenditures; (b) real business fixed investment; (c) real GDP; (d) inflation, measured by the log difference of the GDP price deflator; (e) the *AI Economy Score*; (f) the quarterly value-weighted excess stock market return; (g) the nominal ten-year Treasury yield; and (h) the nominal effective federal funds rate. Consumption, investment, and GDP enter in log differences, and each variable enters with two lags. We identify the shock recursively, placing the score where Gilchrist and Zakrajšek (2012) place the excess bond premium. Real activity and prices therefore respond to a change in managerial expectations only with a lag, while financial variables, being forward-looking, can respond within the quarter.

Figure 4 presents the impulse response functions of the endogenous variables in response to an orthogonalized shock to managerial expectations, as measured by the *AI Economy Score*. A one-standard-deviation shock to the *AI Economy Score* raises consumption, investment, and output over the following 20 quarters, and the magnitudes are large. The accumulated responses of consumption and output growth each peak at roughly 2.5 percentage points above trend six to eight quarters out, then settle near 2 percentage points. Investment responds far more strongly, peaking at roughly 8.5 percentage points about eight quarters after the shock and still running near 6 percentage points at the end of the horizon, consistent with fixed investment being the most volatile component of GDP. The economic surge is accompanied by a transitory rise in inflation, which peaks around 0.2 percentage points five to six quarters after the shock before dissipating and becoming statistically indistinguishable from zero. The expectation shock itself is short-lived: the *AI Economy Score* reverts to its pre-shock level within roughly six quarters.

After a shock to managerial economic forecasts, monetary policy tightens: the federal funds rate rises immediately, peaks at about 0.3 percentage points within the first year, and remains elevated for roughly 16 quarters before converging back toward zero. The ten-year Treasury yield also rises on impact, by about 0.15 percentage points, though this response fades over the following eight quarters and only partially re-emerges later in the horizon. The stock market responds strongly, with the accumulated excess market return climbing to about 4 percentage points within the first year of the shock and remaining positive throughout the horizon. In sum, differences in the *AI Economy Score* orthogonal to current conditions predict substantial subsequent movements in a range of economic variables.

[Insert Figure 4 Here]

### 4.4. AI Forecasts of Additional Macro Variables

We also analyze three other macro variables: (i) aggregate production, (ii) aggregate employment, and (iii) aggregate wages. All three are included in both the Duke CFO Survey and the industry-level data available in the BEA database. The corresponding AI scores are three of the thirteen additional topics described in Section 6.3, namely production quantity, number of employees, and wages and salaries, produced by the same two-pass procedure as the *AI Economy Score* and on the same seven-point scale. In unreported results, we find that the three AI scores significantly predict real outcomes, even after controlling for *Term Spread*, *Real FFR*, and *GZ Spread*, in a setup similar to Table 3.

Appendix Figure IA.1 presents the time series of the three AI scores alongside aggregate production, nonfarm payroll employment, and aggregate wages, where the real outcomes are measured as the percent change in quarter $t+1$ relative to quarter $t-3$. In the graphs, the AI conference transcript-based scores closely resemble the actual outcomes: both the AI-based scores and the realized series decline during the recessions of 2007–09 and 2020 and rebound sharply in 2021. The 2020 contraction is the deepest in the sample for production and employment, with both series recovering in a pronounced V-shape; the *AI Production Score* also captures the milder industrial slowdown of 2015–16. Again, the AI scores fall concurrently with, rather than ahead of, the COVID-19 collapse, as the pandemic would have been difficult for managers to anticipate. While production and employment are volatile, wages exhibit downward stickiness: aggregate wage growth remains positive even during recessions. Wage growth does, however, accelerate substantially in the post-pandemic period, peaking in 2022 before moderating through 2025, and the *AI Wage Score* registers sharp but short-lived drops in 2008 and 2020.

Appendix Figure IA.2 illustrates the variations in the three AI scores across different industries. Darker red shades indicate periods of anticipated contractions, whereas blue shades signify expansions. The impact varies across industries. During the 2007–09 crisis, the *AI Production Score* and *AI Employment Score* dropped most sharply in the Wholesale Trade and Transportation and Warehousing sectors, with weakness in employment scores also extending to Retail Trade, Construction, and Manufacturing. The Mining, Oil and Gas sector displays depressed employment scores around the oil price collapses of 2015–16 and 2020. Following the COVID-19 recovery, employment scores in 2021–22 rose broadly, most visibly in Health Care and Social Assistance, Accommodation and Food Services, and Transportation and Warehousing. The *AI Wage Scores* are systematically lower and more compressed than the production and employment scores, consistent with the stickiness of wages. Notably, the darker red representing aggregate wage expenses for companies in the Accommodation and Food Services industries in 2020 captures the impact of travel restrictions, and renewed red shading across the trade and transportation sectors appears in 2022.

## 5. AI Forecasts on Economic Indicators at Micro Levels

Since our national-level *AI Economy Score* is aggregated from firm-level managerial measures, we can obtain predictors at more disaggregated levels, such as the industry-level and firm-level. In this section, we analyze the predictive power of managerial forecasts for the US economy at the micro level.

Our measures in this section are all based on the score from the question about the outlook for the US economy, observed at the firm-quarter level. The national score in Section 4 is its average across firms, and the industry score is its average across firms in a sector. Since the three levels are nested averages of the same variable, entering them in raw form puts three highly correlated regressors in one model, which inflates the standard errors and leaves the individual coefficients difficult to interpret. We use an orthogonal decomposition instead. Let $s_{i,t}$ be the US-economy score of firm $i$ in period $t$, let $\bar{s}_{j,t}$ be its equally-weighted average across the firms in sector $j$, and let $\bar{s}_t$ be its equally-weighted average across all firms. We define

$$S_t = \bar{s}_t, \qquad D^{ind}_{j,t} = \bar{s}_{j,t} - \bar{s}_t, \qquad D^{firm}_{i,t} = s_{i,t} - \bar{s}_{j,t}. \tag{2}$$

We continue to call $S_t$ the *AI Economy Score*, $D^{ind}_{j,t}$ the *AI Industry Score*, and $D^{firm}_{i,t}$ the *AI Firm Score*. The last two are adjusted scores, measured net of the national average and net of the sector average respectively. The *AI Firm Score* is the economy question answered at the firm level rather than a forecast of the firm's own prospects: it records how one manager reads the US economy relative to managers in the same sector. Because the three components are orthogonal by construction, their coefficients are separately identified. $S_t$ picks up the common movement in managerial expectations, $D^{ind}_{j,t}$ the departure of a sector from that common view, and $D^{firm}_{i,t}$ the departure of a manager from sector peers. No information is lost, since the three sum back to $s_{i,t}$. The industry-level analyses use the first two components and the firm-level analyses use all three.

One implication matters for reading the tables. The adjusted scores hold no common time-series variation, so a specification that includes them without the national component has a lower $R^2$ than the corresponding raw-score specification. Those $R^2$ values are not comparable to the columns that include $S_t$ or the *GZ Spread*.

### 5.1. AI Industry Score and Industrial Economic Indicators

We classify firms into 19 sectors based on the NAICS industrial classification standard, then convert the firm-level AI-predicted scores to industry-level scores by calculating the equally-weighted average within each sector. We use the AI-predicted score for the US economy as our baseline, similar to the national-level analysis. Panel A in Table 5 presents the summary statistics of the main variables used at the industry level. Among the predicted variables, the industrial employment and wage data are only available annually. Therefore, we analyze employment and wage data on an annual basis. We construct annual AI-predicted scores by averaging the quarterly scores for each year. Similarly, the annual *Term Spread* and *Real FFR* are obtained by averaging the monthly measures for each year. The adjusted industry score is a within-period deviation and so is centered near zero, but its dispersion is close to that of the national score: a standard deviation of 0.457 against 0.427 quarterly, and 0.415 against 0.393 annually.

[Insert Table 5 Here]

Table 6 presents the coefficient estimates of various predictors for real sectoral GDP. Panel A displays the short-term results. The improvement in goodness of fit brought by the *AI Economy Score* matches that of the *GZ Spread*, as shown in columns (1) and (2). In column (3), the *AI Industry Score* predicts industrial real GDP on its own, with a coefficient of 0.0122 and a $t$-statistic of 3.93. Its $R^2$ is much lower than in columns (1) and (2) because the adjusted score is identified from differences across sectors within a period and carries none of the aggregate time-series variation. Column (4) adds both AI components to the existing predictors. The national score enters at 0.0199 ($t$ = 2.28) and the industry score at 0.0116 ($t$ = 4.09), and both survive the inclusion of the *GZ Spread*. The magnitudes are economically meaningful: a one-standard-deviation increase in the *AI Economy Score* (0.427 points) raises next-quarter sectoral real GDP growth by 0.85 percentage points, and a one-standard-deviation increase in the *AI Industry Score* (0.457 points) raises it by 0.53 percentage points. Panel B shows that the *AI Economy Score* predicts industrial real GDP through the fourth quarter and again at 16 quarters, while the *AI Industry Score* is significant at the 1% level at every horizon out to 16 quarters. The *GZ Spread* is significant only from the third through the twelfth quarter.

[Insert Table 6 Here]

Table 7 extends the analysis to other economic indicators at the industry level. The *AI Economy Score* in column (2) has predictive power comparable to the *GZ Spread* in column (1) for both employment and wages. On its own in column (3), the *AI Industry Score* is positive and significant at the 1% level for both outcomes, entering at 0.0396 ($t$ = 2.86) for employment and 0.0518 ($t$ = 3.51) for wages. It stays significant in column (4), at 0.0279 ($t$ = 2.28) and 0.0397 ($t$ = 3.16) respectively, whereas the national component turns insignificant in both panels once the *GZ Spread* and the sectoral deviation are included. A one-standard-deviation increase in the adjusted industry score (0.415 points) corresponds to 1.16 percentage points of annual employment growth and 1.65 percentage points of annual wage growth. The sector-specific part of managerial expectations therefore carries information about labor outcomes on its own, and it is the component that survives when all the predictors are added jointly. Table IA.1 of the Internet Appendix reports long horizon results, where the *AI Industry Score* predicts industrial employment and wages for up to four years and the *AI Economy Score* is significant only for wages at three years.

[Insert Table 7 Here]

### 5.2. AI Firm Score and Firm Financial Indicators

To further leverage the micro-level predictive power of our measure, we conduct the analysis at the most granular level, focusing on predicting firm-level outcomes. Panel B in Table 5 presents the summary statistics of firm-level variables. The measure is the same throughout, the manager's outlook for the US economy, observed at the firm level rather than averaged. The *AI Firm Score* is that firm-level score net of its sector average, so it isolates the part of a manager's read on the economy that is specific to the firm rather than shared with comparable firms. Note that firm-level employment and wages are only available on an annual basis. Since the firm-level *Wage* variable (xlr or "Staff Expense - Total") in Compustat has limited availability, the sample size is significantly reduced for the firm-level wage analysis.

Table 8 presents the results for predicting value-added (a close analogue to GDP at the firm level), sales, employment, and wages respectively from Panel A to Panel D. The three AI-based variables enter as the orthogonal components defined in equation (2). In columns (1) to (4), we include only one predictor at a time while controlling for other national-level and firm-level predictors, allowing us to compare their predictive power. Throughout all five columns of the specifications, we control for *Size*, *Book-to-Market*, and *Tangibility*. Firm-fixed effects are included in all specifications. Note that the number of observations may vary in different panels depending on the availability of target and control variables.

[Insert Table 8 Here]

All three AI-based components carry independent predictive power. In column (5) of Panel A, the *AI Economy Score* enters at 0.0857 ($t = 2.20$), the *AI Industry Score* at 0.0772 ($t = 6.80$), and the *AI Firm Score* at 0.0268 ($t = 15.15$) for value-added over the next quarter. Sales in Panel B follow the same pattern, though the national component is significant only at the 10% level once the other predictors are included. The industry and firm components barely move between the single-score columns and column (5), from 0.0802 to 0.0772 and from 0.0274 to 0.0268 for value-added, which is the payoff of the decomposition: each is identified from its own variation rather than competing for the same variation. The national component does attenuate, from 0.104 to 0.0857 for value-added and from 0.0721 to 0.0507 for sales, because column (5) also adds the *GZ Spread*, which is itself an aggregate time series. Evaluated at one standard deviation of each component, the column (5) estimates imply 3.09, 1.84 and 2.85 percentage points of next-quarter value-added growth for the national, industry and firm components respectively, and 1.83, 0.64 and 1.90 percentage points of sales growth.

The annual panels tell the same story, with the national component the weaker of the three. For employment in Panel C, all three are significant in column (5): the *AI Economy Score* at 0.0780 ($t = 2.58$), the *AI Industry Score* at 0.0494 ($t = 3.49$), and the *AI Firm Score* at 0.0425 ($t = 9.60$). For wages in Panel D the national component is insignificant both on its own and jointly ($t = 1.62$ and $t = 0.21$), while the industry and firm components are significant at the 1% level throughout, entering at 0.0683 ($t = 3.26$) and 0.0364 ($t = 4.74$) in column (5). The *AI Firm Score* is significant at the 1% level in every column in which it appears, in every panel, and carries the highest $t$-statistic in each; unlike the two aggregate components, it gains rather than loses precision when the others are included. A one-standard-deviation increase in the firm component corresponds to 3.41 percentage points of annual employment growth and 2.82 percentage points of annual wage growth, against 1.14 and 1.90 percentage points for the industry component. Since the firm component is the manager's view of the economy net of the sector average, the idiosyncratic part of a manager's macroeconomic outlook is what carries information about that firm's own future activity.

Table IA.2 of the Internet Appendix reports the long-horizon results at the firm level. The *AI Firm Score* is positive and significant at the 1% level for value-added and sales at every horizon out to 16 quarters, and for employment out to four years, controlling for known macroeconomic indicators and firm covariates. Wages are the exception: the firm component predicts wage growth two years ahead but not at three or four, consistent with wages adjusting slowly.

The firm-level results are further evidence that AI-extracted managerial forecasts carry information about firm prospects, at both short and long horizons, that the known aggregate predictors and firm characteristics do not.

# 6. Dissecting Managerial Forecasts: Behavioral Biases and Multi-dimensional Insights

## 6.1. Heterogeneity in Managerial Behavior Types

Managers differ in their beliefs and styles, which leads to variations in their forecasts for the future economy. This section examines the forecasting performance of different types of managers.

We classify managers along two dimensions. First, to capture forecasting style, we construct a bias measure for each firm-quarter as the difference between the tercile ranking of the manager's AI Economy Score and the tercile ranking of realized real GDP over the same quarter, where real GDP terciles are computed over the full 2006–2025 sample. Averaging this bias measure over each manager's trailing eight quarters and sorting firms into quartiles, we classify a manager as an Optimist if their average bias falls in the top quartile, a Pessimist if it falls in the bottom quartile, and a Realist otherwise. Second, to capture forecasting skill, we classify a manager as Accurate or Inaccurate if the R-squared from regressing real GDP on their firm-level AI Economy Score over the trailing eight quarters falls in the top or bottom quartile, respectively. We then average the AI Economy Score within each of the five groups to construct *AI Economy Score: Accurate Managers*, *AI Economy Score: Inaccurate Managers*, *AI Economy Score: Optimist*, *AI Economy Score: Pessimist*, and *AI Economy Score: Realist*, and use each to predict next-quarter real GDP.

Table 9 reports the results. In column (2), the aggregate AI Economy Score is a strong predictor of next-quarter real GDP, raising the R-squared from 0.440 to 0.487. When we split by skill, the score constructed from Accurate managers alone remains highly significant in column (3), while the score from Inaccurate managers in column (4) is statistically indistinguishable from zero. This pattern holds when both are included jointly in column (5): the Accurate managers' score retains its significance, while the Inaccurate managers' score remains insignificant. When split by style, all three types of managers, Pessimists, Optimists, and Realists, strongly predict future GDP. The score based on Realists delivers the largest coefficient and the highest single predictor R-squared of 0.489, and it remains significant when all three style based scores are included together in the regression in column (9).

Taken together, these results suggest that the predictive power of managerial forecasts for the real economy is concentrated among managers who are historically accurate or unbiased (Realists), whereas forecasts from historically inaccurate managers, and to a lesser extent from persistently optimistic or pessimistic managers, contain little incremental predictive content. This evidence is more consistent with an information-based channel than with a sentiment- or noise-driven channel: AI-extracted managerial forecasts appear to predict real economic activity primarily because skilled managers possess and transmit superior information about future fundamentals, rather than because managerial beliefs per se exert an independent effect on real outcomes. This evidence also relates to the *news* vs *noise* debate (e.g., Barsky and Sims, 2011; Chahrour and Jurado, 2018; Schmitt-Grohé and Uribe, 2012; Blanchard, L'Huillier, and Lorenzoni, 2013) in the macroeconomics literature as our results is more supportive of the news or information channel rather than a noise channel.

[Insert Table 9 Here]

### 6.2. Heterogeneity in Business Diversification

Managers in firms with varying levels of concentration may differ in their ability to forecast national economic conditions. One possibility is that managers in more diversified firms have access to a broader set of information channels, as they must engage with various sectors of the economy. This increased exposure could lead to more informationally-relevant economic forecasts. However, prior literature suggests that more diversified firms tend to underperform compared to their more focused counterparts due to higher agency costs and fewer growth opportunities (Lang and Stulz, 1994; Berger and Ofek, 1995). If this finding extend to information aggregation, managers in diversified firms may generate forecasts that are less accurate than the focused firms with better understanding of their operations.

To test these two opposing hypotheses, we investigate the relationship between a firm's level of concentration and the predictive power of future economic conditions. We measure concentration level using two proxies. First, we employ the business operation Herfindahl-Hirschman Index (HHI), calculated as the sum of the squared ratio of each business segment's sales to total sales. *High Business Segment HHI* is defined as an indicator variable that takes the value of one if the Business Segment HHI exceeds the median for all firms within a quarter. Second, we use GEO HHI, which is the sum of the squared ratios of the frequency with which each state is mentioned in a firm's 10-K filing relative to the number of different states mentioned (Garcia and Norli, 2012). *High GEO HHI* is an indicator variable that equals one if the GEO HHI is above the median for all firms in a given quarter.

Table 10 presents the results. In columns (1) to (4), we report the interactions between the *AI Economy Score* and the diversification measures in firm-level panel regressions. All specifications control for the four lags of the dependent variable, other macroeconomic predictors, and firm fixed effects. The two interaction terms in columns (1) and (2) consistently exhibit positive and statistically significant coefficients, indicating that the predictive power of economic forecasts is stronger for more focused firms than for more diversified firms. The two interaction terms in columns (3) and (4) show positive coefficients, but they are not significant. These results support the hypothesis that economic forecasts from more focused firms have more predictive power, possibly due to their deeper knowledge of their industry.

[Insert Table 10 Here]

### 6.3. Integrating Multi-dimensional Information from Managerial Forecasts

Managerial discussions in conference calls contain a rich set of information, including expectations about multiple aspects of firms' performance, operations, and policies, industry trends, and economic conditions. These discussions may contain additional implicit information about future economic growth beyond what is contained in explicit managerial economic forecasts. To capture such information, we also extract managers' forecasts regarding changes in various economic and firm-related aspects for the next quarter. In addition to asking about their expectations for optimism in the U.S. economy, we prompt ChatGPT with thirteen other questions covering managers' expectations regarding changes in optimism about the global economy, their firm's financial prospects, their industry's financial prospects, earnings, revenue, investments, wages and salaries, number of employees, demand for their products or services, production quantity, product or service prices, input or commodity prices, and the cost of capital.

These thirteen aspects are extracted and scored with a similar two-pass procedure as that used for the US economy question, on the same seven-point scale, with all thirteen scored in a single request per conference call. In the first pass, we extract manager's statements relevant to the thirteen questions, respectively. We then obtain scores based on the first-pass outputs in the second pass.[6] Based on the responses from ChatGPT, we first obtain firm-level AI scores for each prompted aspect and then construct national-level scores by averaging firm-level scores.

[6] The prompts and pipleines we use are reported in the Internet Appendix.

To synthesize the information contained in these scores, we compose a new measure, *AI Weighted Score*, that is a linear combination of the 14 scores as follows. First, for each quarter $t$, we conduct the following panel regression, estimated without an intercept (i.e., through the origin) via ridge regression with a negligible penalty ($\lambda = 0.001$, included only for numerical stability and imposing effectively no shrinkage relative to OLS):

$$Sales_{i,s} = \sum_{k=1}^{14} \beta_{k,t} AI\ Score_{k,i,s-1} + \varepsilon_{i,s}, \quad s \leq t-1. \tag{3}$$

where $AI\ Score_{k,i,s-1}$ is the firm-level AI Score for the $k$-th aspect of firm $i$ in quarter $s-1$. In other words, we forecast firm sales via a multilinear regression on the 14 firm-level scores in the previous quarter. The regression uses the entire set of firm-quarter observations available prior to quarter $t$ so that the coefficients ($\beta_{k,t}$) are estimated on the full expanding history and are known at quarter $t$ without look-ahead bias. We then compose the weighted score at the national level:

$$AI\ Weighted\ Score_t = \sum_{k=1}^{14} \beta_{k,t} AI\ Score_{k,t}, \tag{4}$$

where $AI\ Score_{k,t}$ is the $k$-th national AI score in quarter $t$. We then use this AI weighted score to predict economic growth at the national level.

Table 11 presents the national-level results. Relative to the *AI Economy Score* baseline (Table 3), the *AI Weighted Score* substantially extends the persistence of GDP predictability: it remains statistically significant through the 8th quarter – the longest horizon reported – whereas the *AI Economy Score* loses significance after the 5th quarter ($t = 2.21$ at $h = 5$, falling to $t = 0.72$ at $h = 6$). The pattern in fit, however, is a trade-off rather than a uniform improvement: at short horizons ($h = 1$ through $h = 5$) the *AI Economy Score* actually achieves higher $R^2$ than the *AI Weighted Score* (e.g., 0.486 vs. 0.465 at $h = 1$; 0.566 vs. 0.522 at $h = 3$), but the *AI Weighted Score*'s $R^2$ overtakes the baseline from $h = 6$ onward (0.569 vs. 0.565 at $h = 6$) and continues rising through $h = 8$ (0.620), a horizon at which the baseline is no longer estimable with any precision. In short, combining the 14 dimensions via estimated weights sacrifices some near-term explanatory power in exchange for a materially longer and more stable predictive horizon.

[Insert Table 11 Here]

Drawing on a broader range of topics managers discuss therefore extends the predictive horizon for real GDP growth and strengthens the power especially in the long run.

## 7. Robustness Tests and Additional Analyses

This section conducts robustness tests of the previous results.

### 7.1. Horse Race with Survey Forecasts

We conduct a horse race between our *AI Economy Score* and the forecasts of real GDP growth by professional forecasters in the SPF. The SPF is a survey that provides economic forecasts and has been widely used by researchers for macroeconomic expectations (e.g., Coibion and Gorodnichenko, 2012, Coibion and Gorodnichenko, 2015, Bordalo, Gennaioli, and Shleifer, 2020). It contains quarterly forecasts of multiple economic indicators, including real GDP. For the survey conducted at quarter $t$, we take the median forecasts of real GDP by all the respondents for quarter $t+1$, then construct the forecasted GDP growth rate using the realized real GDP in quarter $t-1$ as the base, because it is the most recent available realized GDP number when economists make forecasts.

The results are presented in Table 12. The dependent variable is the log difference of real GDP for the next quarter, consistent with our baseline specification. In column (1), we include the *Term Spread*, *Real FFR*, *GZ Spread*, and the *AI Economy Score*. In column (2), we replace the *AI Economy Score* with the SPF-Forecasted real GDP growth. We include both measures simultaneously in column (3). The $R^2$ comparison between columns (1) and (2) shows that the in-sample fit for the model including the *AI Economy Score* (0.486) is comparable to that of the model including the *SPF-Forecasted rGDP* (0.491), with the survey-based measure only marginally ahead. When we include both measures in the model in column (3), neither loses its predictive power: the *AI Economy Score* (0.0110) and the *SPF-Forecasted rGDP* (0.249) both remain significant at the 10% level, though each coefficient is roughly half its size relative to the single-measure specifications. This result indicates that the *AI Economy Score* is a predictor of real GDP for the near future that is comparable in strength to survey-based forecasts, and that the two measures likely contain at least partially independent information.

[Insert Table 12 Here]

We also re-estimate the vector autoregression of Section 4.3 with the SPF GDP forecast added as an endogenous variable, to see whether the *AI Economy Score* carries information beyond what professional forecasters already publish. Appendix Figure IA.3 reports the impulse responses; every panel of the baseline figure survives the addition. The SPF forecast itself rises after a positive shock to the score and stays significantly above zero well past the first year, so the score moves professional forecasts as well as realized outcomes.

### 7.2. Anonymization Tests

The scoring prompt supplies the call date and instructs the model to reason using only what it would have known at that date, but the model's training data extend well past the dates of most calls in our sample, so one potential concern is that in answering queries about conference calls the model could also have drawn on information unavailable at the time of the calls. To address this potential "look-ahead bias," we re-run our results after removing all identifying information in the conference call transcripts, including firm information and dates. Glasserman and Lin (2023) demonstrate that look-ahead bias can be mitigated by removing company names from the prompt's text. We employ a two-step process to anonymize firm-specific text. In the first step, we follow the procedure outlined in Engelberg et al. (2025) and paraphrase the conference call text one paragraph at a time using the Gemma-4B model. The model is instructed to fully anonymize each paragraph by replacing all identifiable details, including firm names, tickers, industries, products, numerical values, and dates, with generic placeholders so that neither the company nor its sector can be inferred (see the full prompt in Internet Appendix: Anonymization Prompts). We apply this procedure to a random 15% sample of the data. The Gemma-4B model is chosen for this step because its token-level computation cost is lower than that of larger LLMs, which is particularly important given that the anonymization must be performed at the paragraph level.

In the second step, we further anonymize the paraphrased conference call texts by removing any remaining dates, personal names, organization names, and product names that might persist after the initial paraphrasing. Specifically, we use regular expressions to mask all years (ranging from 1900 to 2099) and month names, including their common abbreviations. We then employ Python's spaCy library (using the en_core_web_sm model model) to identify and redact entity names related to persons, organizations, and products, ensuring comprehensive anonymization of the text.

Once we obtain the anonymized conference call transcripts, we prompt ChatGPT-5-nano to de-anonymize each text by inferring the corresponding company, industry, and year based on contextual clues (see the full prompt in Internet Appendix: Anonymization Prompts). We intentionally use the smallest OpenAI model to minimize its ability to correctly identify the underlying firm or period, ensuring that any remaining signal in the text reflects economic content rather than identifiable metadata. The ChatGPT-5-nano model correctly identified the company associated with the earnings call in 34.7% of the anonymized conference call transcripts. We therefore focus on the remaining 65.3%—which corresponds to about 10% of the total sample in our baseline analysis (15% × 65.3%)—to re-estimate our baseline model. Using these texts and the GPT-5-nano model, we follow our previous procedure to construct the variable *AI Economy Score_Entity Neutered*.

The results, reported in Table 13, indicate that *AI Economy Score_Entity Neutered* remains positively and significantly associated with real GDP growth in the subsequent quarter, even after controlling for established predictors of GDP growth. The estimated coefficients of determination are similar in magnitude to those presented in Table 3. This mitigates the look-ahead concern: the score draws on the language of the calls rather than on identifying details such as firm names or dates.

[Insert Table 13 Here]

### 7.3. Out-of-sample Tests

In this section, we further address the concern that the AI model may utilize public information beyond the content of a specific earnings call, by conducting completely out-of-sample tests. To mitigate this concern, we conduct our analysis on a subsample of conference calls from 2024 Q2 to 2025 Q4, a period that falls outside the GPT-5-mini model's training window. Since the GPT-5-mini model cannot access events or information beyond its training cutoff, any forecasting power obtained from this subsample cannot be attributed to look-ahead bias. Because the macro-level sample in this window is too short for reliable regression estimation at the national level, we instead focus on the firm- and industry-level analyses. Specifically, we examine whether the *AI Firm Score*, together with the *AI Economy Score* and the *AI Industry Score*, can forecast firm-level *Sales* and industry-level *Real GDP* for the next quarter, using the same orthogonal decomposition as in the main tables.

Table 14 shows that the results are strongest for the *AI Firm Score*. In Panel A, it predicts next-quarter firm-level sales both alone (column (4)) and jointly with the other components and the *GZ Spread* (column (5)). The *AI Economy Score* and the *AI Industry Score* are insignificant in this panel, as is the *GZ Spread*. Panel B shows a similar pattern at the industry level, where the *AI Industry Score* predicts industry Real GDP both alone (column (3)) and jointly with the other predictors (column (4)) while the *AI Economy Score* is insignificant in both. These results indicate that the predictive power of the *AI Firm Score* and the *AI Industry Score* is not driven by look-ahead bias, though the evidence is weaker for the more aggregated *AI Economy Score* in this subsample.

[Insert Table 14 Here]

### 7.4. Alternative Generative AI Model

In this section, we apply the same methodology described in 3.1, but instead of using the GPT-5 mini model, we implement an alternative open-source language model, Gemma, developed by Google. Being fully open-source, Gemma can be run locally at no cost, making it highly accessible for academic research.

Table 15 shows that the results obtained from Gemma are broadly consistent with those from GPT-5 mini. Panel A reports the one-quarter-ahead results and Panel B reports longer horizons. In Panel A, column (2), the coefficient on *AI Economy Score_Gemma* remains significant at the 1% level, although the t-statistic (6.26) and R-squared (0.425) are slightly lower than their counterparts in 3 (6.86 and 0.471). In Panel B, the predictive power remains significant up to three quarters ahead, compared with five quarters in 3. The findings are therefore robust to the choice of language model.

[Insert Table 15 Here]

### 7.5. Controlling for Corporate Loan Spreads

Given prior evidence that corporate loan spreads, constructed from secondary market prices, forecast business cycles and outperform other credit-spread measures (Saunders et al., 2025), we examine whether managerial communications convey information beyond what is already incorporated into loan spreads.

We include the *Loan Spread* as a control variable in our macroeconomic forecasting regressions. As shown in Table 16, the *AI Economy Score* remains a statistically significant predictor of GDP growth at all horizons up to six quarters ahead, even after controlling for loan spreads. These findings suggest that the *AI Economy Score* captures incremental information not embedded in traditional credit market indicators.

[Insert Table 16 Here]

### 7.6. Controlling for Sentiment of Conference calls

Given that manager sentiment has been shown to predict financial market movements (Jiang et al., 2019), a potential concern is that the AI Economy Score may primarily capture manager sentiment, or be highly correlated with it—information that can be extracted using a standard bag-of-words approach. To address this concern, we construct a measure of manager sentiment based on the Loughran-McDonald dictionary. Specifically, we define *Sentiment* as the difference between the number of positive and negative words, divided by the total number of positive and negative words in each earnings call. We then aggregate this measure to the macro level by averaging across firms within each quarter.

We include *Sentiment* as a control variable in our macro-level forecasting regressions. The results, reported in Table 17, show that the *AI Economy Score* remains a significant predictor of GDP growth at the 1% level at every horizon we examine, up to six quarters ahead. Notably, the inclusion of manager sentiment does not diminish the predictive power of the AI Economy Score. These findings suggest that the AI Economy Score captures distinct information beyond that embedded in manager sentiment.

[Insert Table 17 Here]

## 8. Concluding Remarks

This paper explores the potentials of generative AI tools, specifically ChatGPT, to extract managerial expectations about the macroeconomy. We propose the *AI Economy Score*, a novel measure capturing the average managerial expectation for the US economy in the next quarter. Our findings demonstrate that the *AI Economy Score* is a strong predictor of future economic activities, including GDP growth, production, employment, and wages, providing additional predictive power beyond those of existing benchmark measures.

The score has three major advantages: wide applicability, long-term predictability, and micro-level insights. First, the *AI Economy Score* can be readily constructed using publicly available data, offering an easily accessible set of expectation variables for researchers, policymakers, and investors. Second, it forecasts a range of macroeconomic indicators up to ten quarters ahead, a horizon useful for long-term decision-making. Third, the same method yields industry- and firm-level scores, and these carry independent predictive power that lasts even longer, up to four years.

Our contribution extends beyond economic forecasting. By uncovering managerial expectations, we gain valuable insights into the thought processes and perspectives of top management regarding the future economic environment. This information can be instrumental for policymakers in crafting effective economic policies and for corporations in formulating strategic business decisions.

Future research should explore the integration of AI-generated forecasts with traditional economic models to enhance forecasting accuracy and examining the applicability of our framework to other economic contexts and regions. As the field of AI continues to evolve, we anticipate further advancements in this area, leading to even more powerful tools for economic analysis and decision-making.

## References

Barsky, R. B., and E. R. Sims. 2011. News shocks and business cycles. *Journal of Monetary Economics* 58:273–89. ISSN 0304-3932. doi:https://doi.org/10.1016/j.jmoneco.2011.03.001.

———. 2012. Information, animal spirits, and the meaning of innovations in consumer confidence. *American Economic Review* 102:1343–77.

Bauer, M. D., C. E. Pflueger, and A. Sunderam. 2024. Perceptions about monetary policy. *The Quarterly Journal of Economics* 139:2227–78.

Ben-David, I., J. R. Graham, and C. R. Harvey. 2013. Managerial miscalibration. *The Quarterly journal of economics* 128:1547–84.

Berger, P. G., and E. Ofek. 1995. Diversification's effect on firm value. *Journal of financial economics* 37:39–65.

Bernanke, B. S., and A. S. Blinder. 1992. The federal funds rate and the channels of monetary transmission. *The American Economic Review* 82:901–21. ISSN 00028282.

Bhandari, A., J. Borovička, and P. Ho. 2022. Survey data and subjective beliefs in business cycle models. *Available at SSRN 2763942* .

Blanchard, O. J., J.-P. L'Huillier, and G. Lorenzoni. 2013. News, noise, and fluctuations: An empirical exploration. *American Economic Review* 103:3045–70.

Bordalo, P., N. Gennaioli, and A. Shleifer. 2020. Memory, attention, and choice. *The Quarterly journal of economics* 135:1399–442.

Bybee, J. L. 2025. The ghost in the machine: Generating beliefs with large language models. *arXiv preprint arXiv:2305.02823* .

Candia, B., O. Coibion, and Y. Gorodnichenko. 2023. The macroeconomic expectations of firms. In *Handbook of Economic Expectations*, 321–53. Elsevier.

Chahrour, R., and K. Jurado. 2018. News or noise? the missing link. *American Economic Review* 108:1702–36.

Coibion, O., and Y. Gorodnichenko. 2012. What can survey forecasts tell us about information rigidities? *Journal of Political Economy* 120:116–59.

———. 2015. Information rigidity and the expectations formation process: A simple framework and new facts. *American Economic Review* 105:2644–78.

Coibion, O., Y. Gorodnichenko, and S. Kumar. 2018. How do firms form their expectations? new survey evidence. *American Economic Review* 108:2671–713.

Coibion, O., Y. Gorodnichenko, and T. Ropele. 2020. Inflation expectations and firm decisions: New causal evidence. *The Quarterly Journal of Economics* 135:165–219.

Coibion, O., Y. Gorodnichenko, and M. Weber. 2022. Monetary policy communications and their effects on household inflation expectations. *Journal of Political Economy* 130:1537–84.

Engelberg, J., A. Manela, W. Mullins, and L. Vulicevic. 2025. Entity neutering Available at SSRN: https://ssrn.com/abstract=5182756 or http://dx.doi.org/10.2139/ssrn.5182756.

Fama, E. F., and K. R. French. 1992. The cross-section of expected stock returns. *the Journal of Finance* 47:427–65.

Garcia, D., and Ø. Norli. 2012. Geographic dispersion and stock returns. *Journal of Financial Economics* 106:547–65.

Gernon, D. 2017. As trump disbanded advisory groups, this is who was in and who was out. https://www.cnbc.com/2017/08/16/heres-whos-in-and-out-of-trumps-economic-advisory-councils.html.

Gilchrist, S., and E. Zakrajšek. 2012. Credit spreads and business cycle fluctuations. *American Economic Review* 102:1692–720.

Glasserman, P., and C. Lin. 2023. Assessing look-ahead bias in stock return predictions generated by GPT sentiment analysis. Working Paper.

Graham, J. R., and C. R. Harvey. 2001. The theory and practice of corporate finance: Evidence from the field. *Journal of financial economics* 60:187–243.

Hansen, A. L., and S. Kazinnik. 2023. Can ChatGPT decipher Fedspeak? *Working paper, Available at SSRN 4399406* .

Jha, M., H. Liu, and A. Manela. 2022. Does finance benefit society? A language embedding approach. *Social Science Research Network. Working paper 3655263* .

Jha, M., J. Qian, M. Weber, and B. Yang. 2023. ChatGPT and corporate policies. *Working paper 32161, National Bureau of Economic Research* .

Jiang, F., J. Lee, X. Martin, and G. Zhou. 2019. Manager sentiment and stock returns. *Journal of Financial Economics* 132:126–49.

Kim, A., M. Muhn, and V. Nikolaev. 2024. Financial statement analysis with large language models. *Chicago Booth Research Paper* .

Kim, A. G., M. Muhn, and V. V. Nikolaev. 2023. Bloated disclosures: Can ChatGPT help investors process information? *Chicago Booth Research Paper* .

Korinek, A. 2023. Generative ai for economic research: Use cases and implications for economists. *Journal of Economic Literature* 61:1281–317.

Lang, L. H., and R. M. Stulz. 1994. Tobin's q, corporate diversification, and firm performance. *Journal of political economy* 102:1248–80.

Li, K., F. Mai, R. Shen, C. Yang, and T. Zhang. 2023. Dissecting corporate culture using generative AI – Insights from analyst reports. *Working paper, Available at SSRN 4558295* .

Lopez-Lira, A., and Y. Tang. 2023. Can ChatGPT forecast stock price movements? return predictability and large language models. *Working paper, University of Florida* .

Newey, W. K., and K. D. West. 1987. Notes and comments: A simple, positive semi-definite, heteroskedasticity and autocorrelation consistent covariance matrix. *Econometrica* 55:703–8.

Ouyang, S., H. Yun, and X. Zheng. 2024. How Ethical Should AI Be? How AI Alignment Shapes the Risk Preferences of LLMs. doi:10.2139/ssrn.4851711.

Pflueger, C., E. Siriwardane, and A. Sunderam. 2020. Financial market risk perceptions and the macroeconomy. *Quarterly Journal of Economics* 1443.

Saeedy, A. 2024. Jamie dimon warns u.s. might see interest-rate spike. https://www.wsj.com/finance/jamie-dimon-warns-u-s-might-face-interest-rate-spike-83789da7.

Saunders, A., A. Spina, S. Steffen, and D. Streitz. 2025. Corporate loan spreads and economic activity. *The Review of Financial Studies* 38:507–46.

Schlingemann, F. P., and R. M. Stulz. 2022. Have exchange-listed firms become less important for the economy? *Journal of Financial Economics* 143:927–58.

Schmitt-Grohé, S., and M. Uribe. 2012. What's news in business cycles. *Econometrica* 80:2733–64.

Sheng, J., Z. Sun, B. Yang, and A. L. Zhang. 2024. Generative AI and asset management. *Review of Financial Studies, accepted* .

van Binsbergen, J. H., S. Bryzgalova, M. Mukhopadhyay, and V. Sharma. 2024. (almost) 200 years of news-based economic sentiment. doi:10.3386/w32026.

Weber, M., F. D'Acunto, Y. Gorodnichenko, and O. Coibion. 2022. The subjective inflation expectations of households and firms: Measurement, determinants, and implications. *Journal of Economic Perspectives* 36:157–84.

Yang, S. 2023. Predictive patentomics: Forecasting innovation success and valuation with ChatGPT. *Working paper, Available at SSRN 4482536* .

**Figure 1.** Explanations for Change in Optimism

We compile all the one-sentence explanations associated with positive (and separately, negative) changes in managerial optimism about the U.S. economy. We then prompt ChatGPT to group these explanations into distinct topics for each type of change.

**(a)** Explanations for Positive Change in Optimism

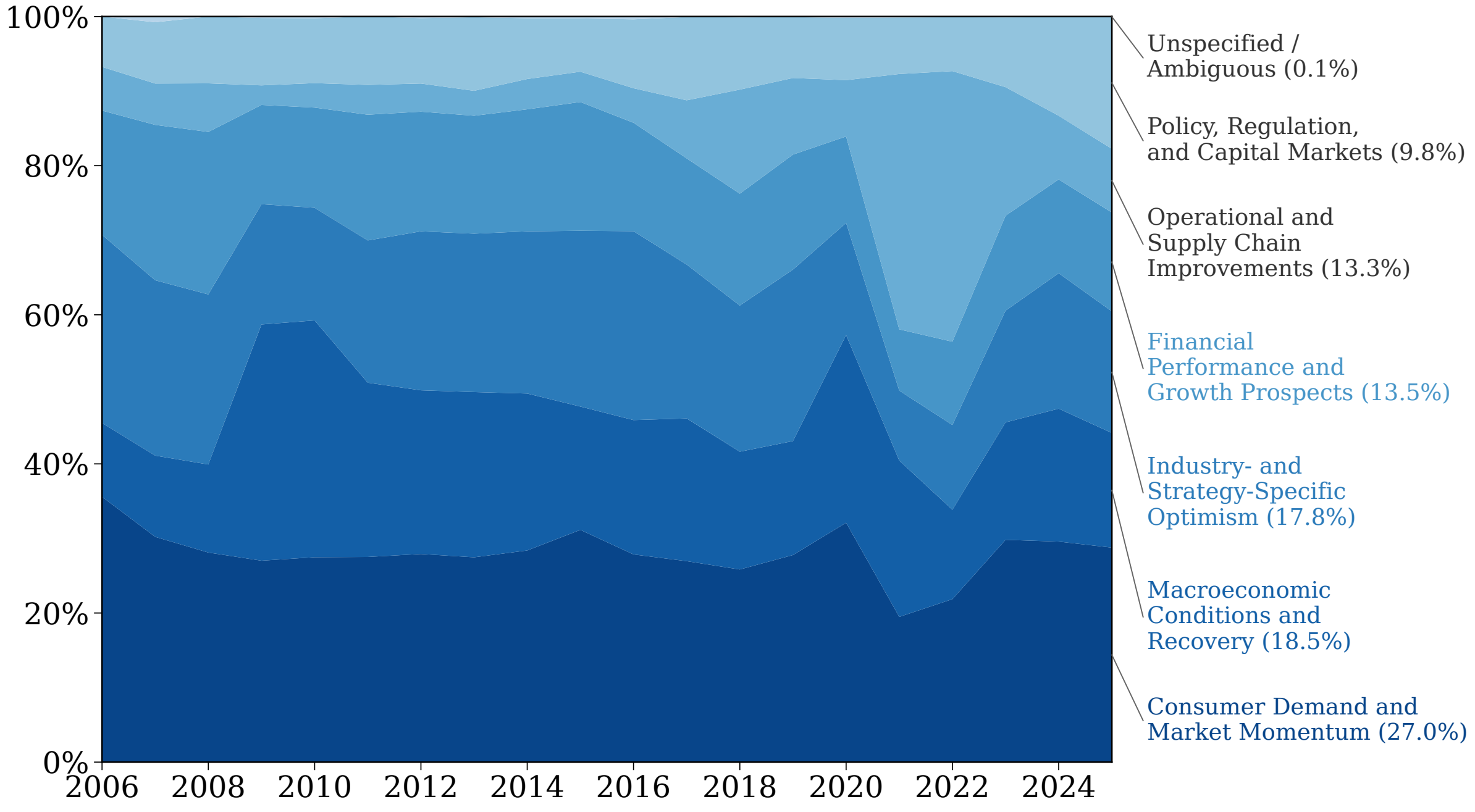

**(b)** Explanations for Negative Change in Optimism

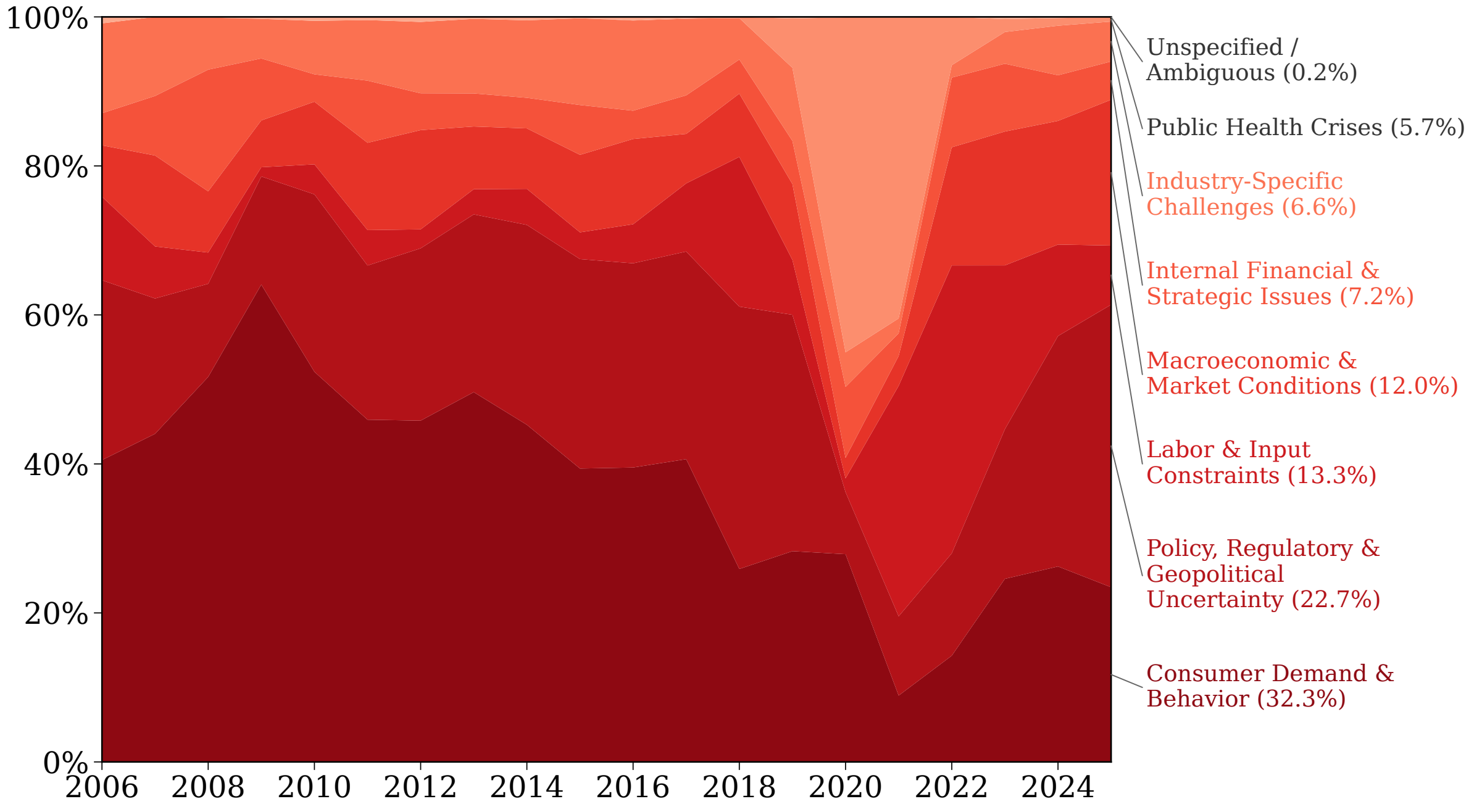

**Figure 2.** AI Economy Score vs. Realized and Forecasted GDP

The AI Economy Score is constructed with the two-pass procedure described in Section 3.1. The first pass extracts from each transcript the statements in which management speaks to conditions in the US economy, and the second pass scores how management expects those conditions to change over the coming quarter, on an integer scale from 1 (severe deterioration) to 7 (strong expansion) with 4 as the neutral midpoint. The firm-quarter scores are aggregated each quarter into an equally weighted nationwide score. The changes in outcomes are (a) the percent change in realized GDP in quarter t+1 relative to quarter t-3, and (b) the change in Survey of Professional Forecasters (SPF)-based GDP forecast for quarter t+1 relative to the realized GDP in quarter t-3.

**(a)** *AI Economy Score* vs. Realized GDP Growth

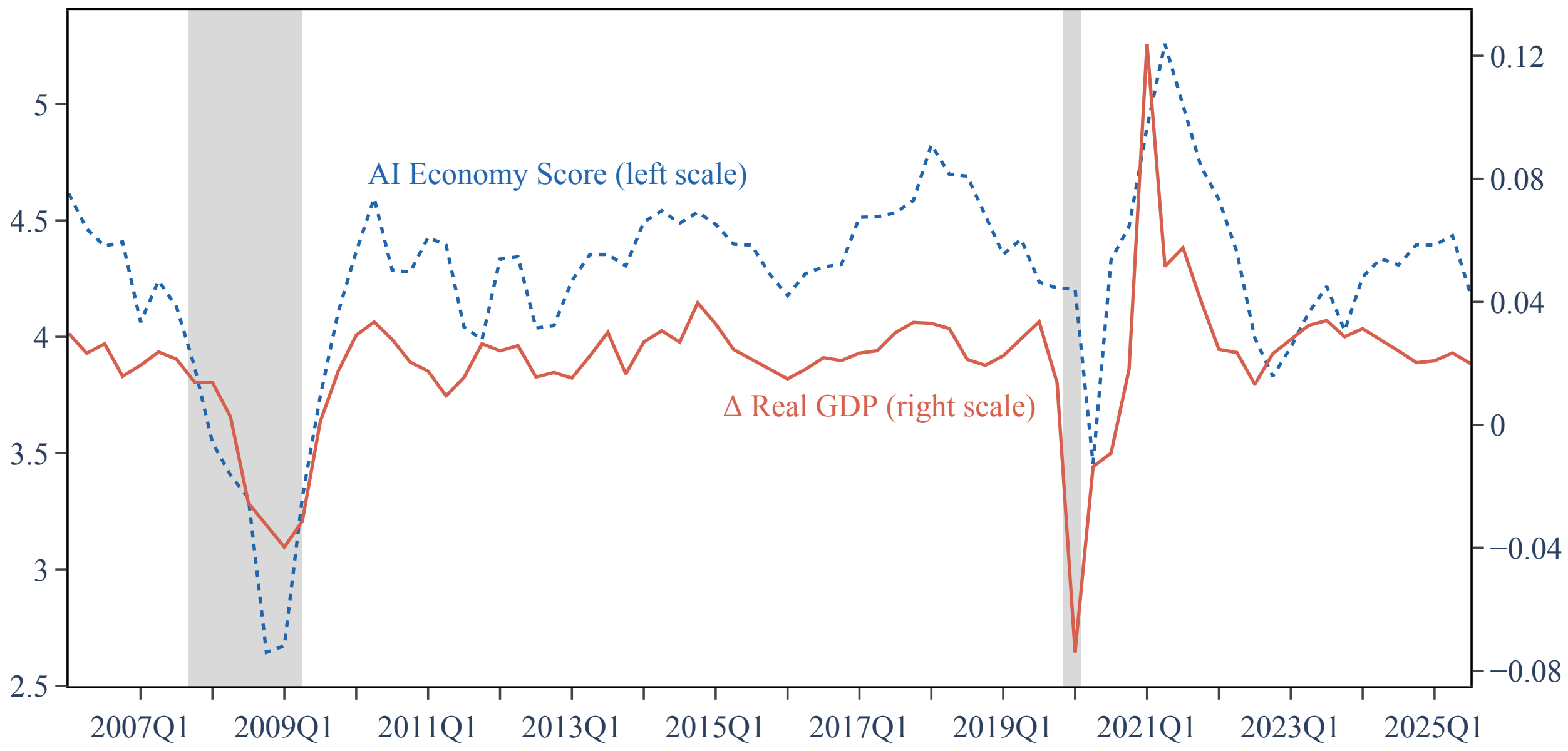

**(b)** *AI Economy Score* vs. SPF Forecasts

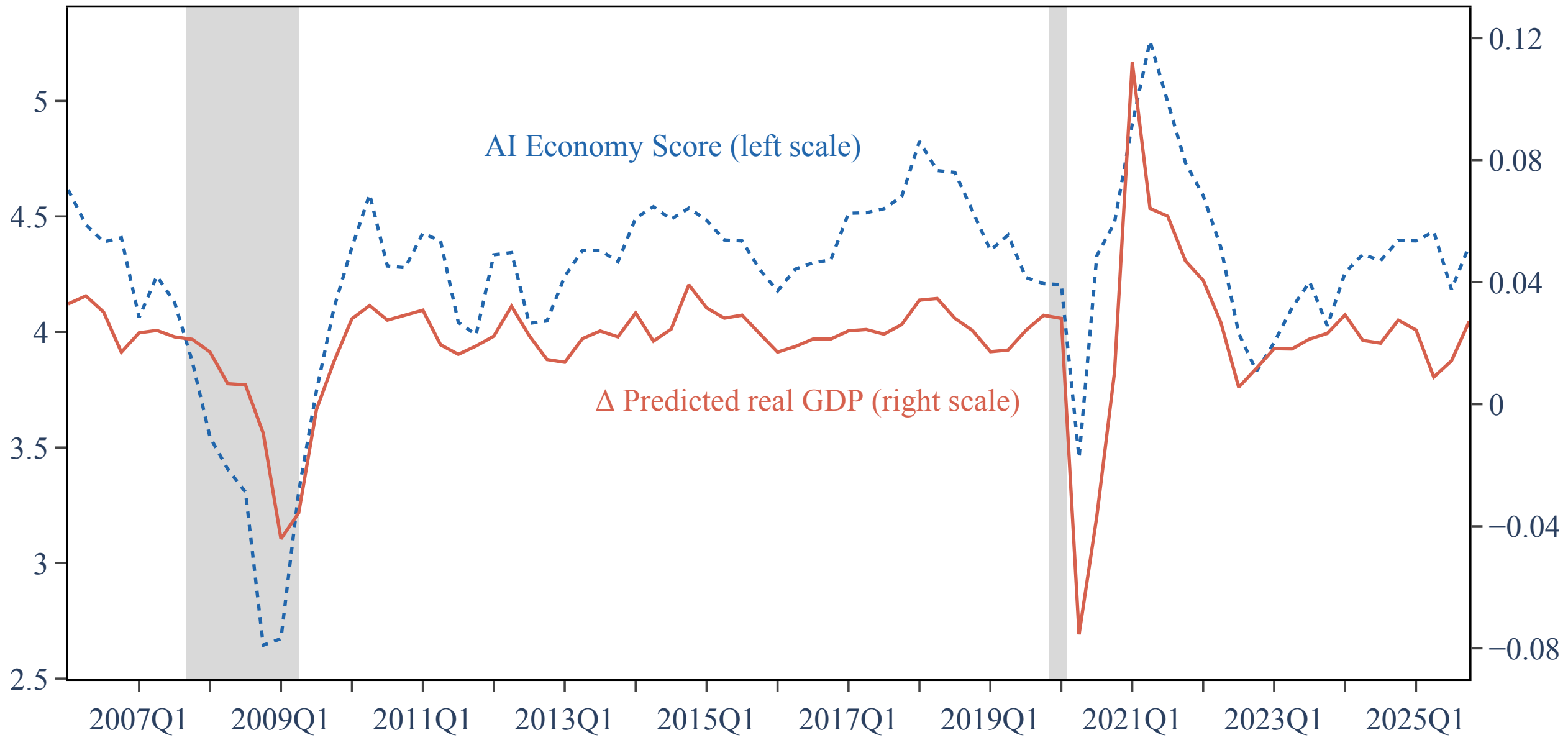

**Figure 3.** *AI Economy Score* across Industries

This figure represents average yearly AI Economy Score across industries. The score is calculated based on conference call texts of the firm (described in Section 3.1). The firms are aggregated into industries, following the North American Industry Classification System (NAICS).

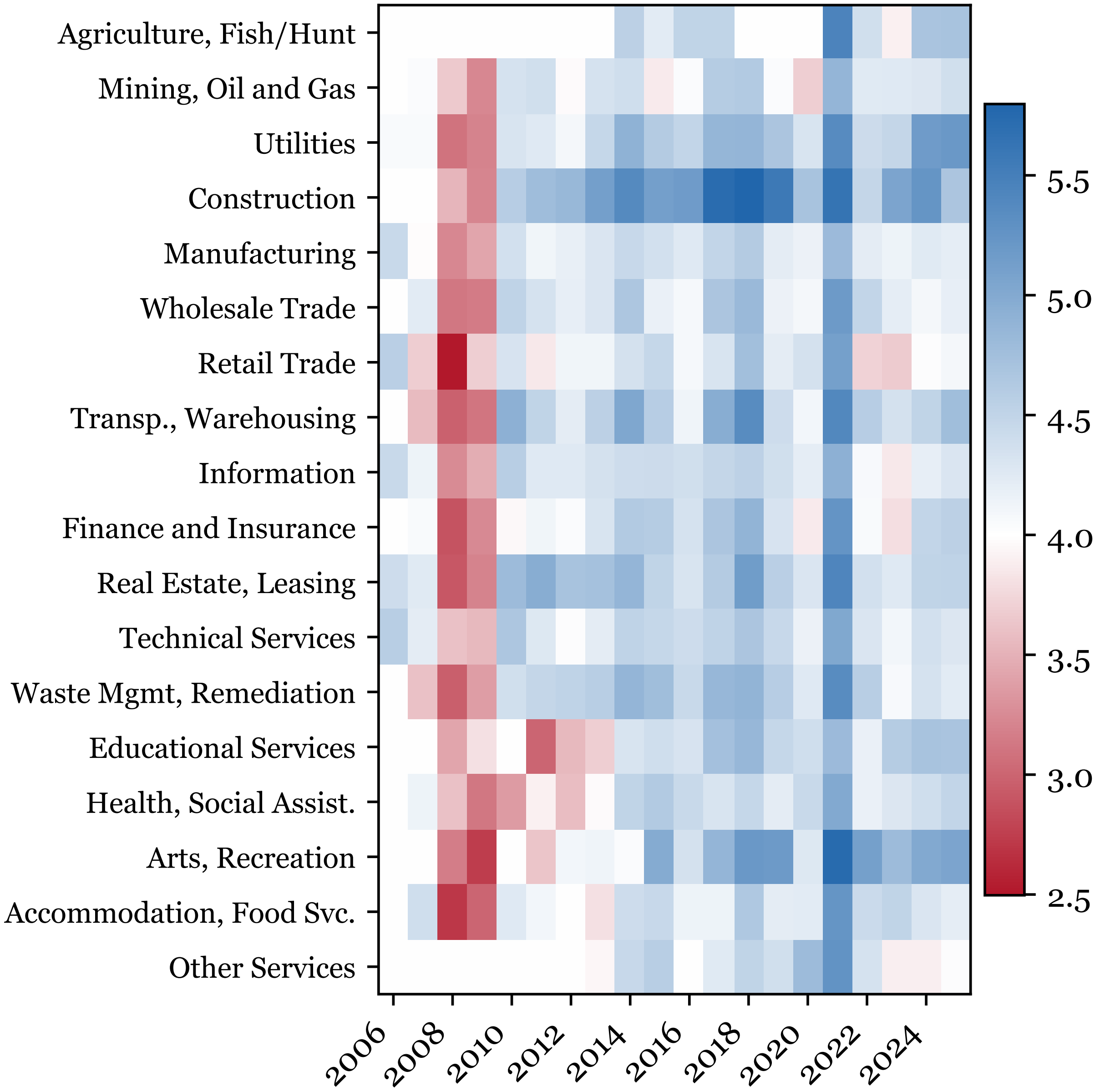

**Figure 4.** Macroeconomic implications of changes in *AI Economy Score*

The figure presents the impulse responses to a one-standard-deviation orthogonal shock to the *AI Economy Score*. The responses of consumption, investment, output growth, and excess market return have been accumulated. Shaded areas represent 95-percent confidence intervals, derived from 2,000 bootstrap replications. In each graph, the X-axis shows the quarters after the shock, and Y-axis is in percentage points.

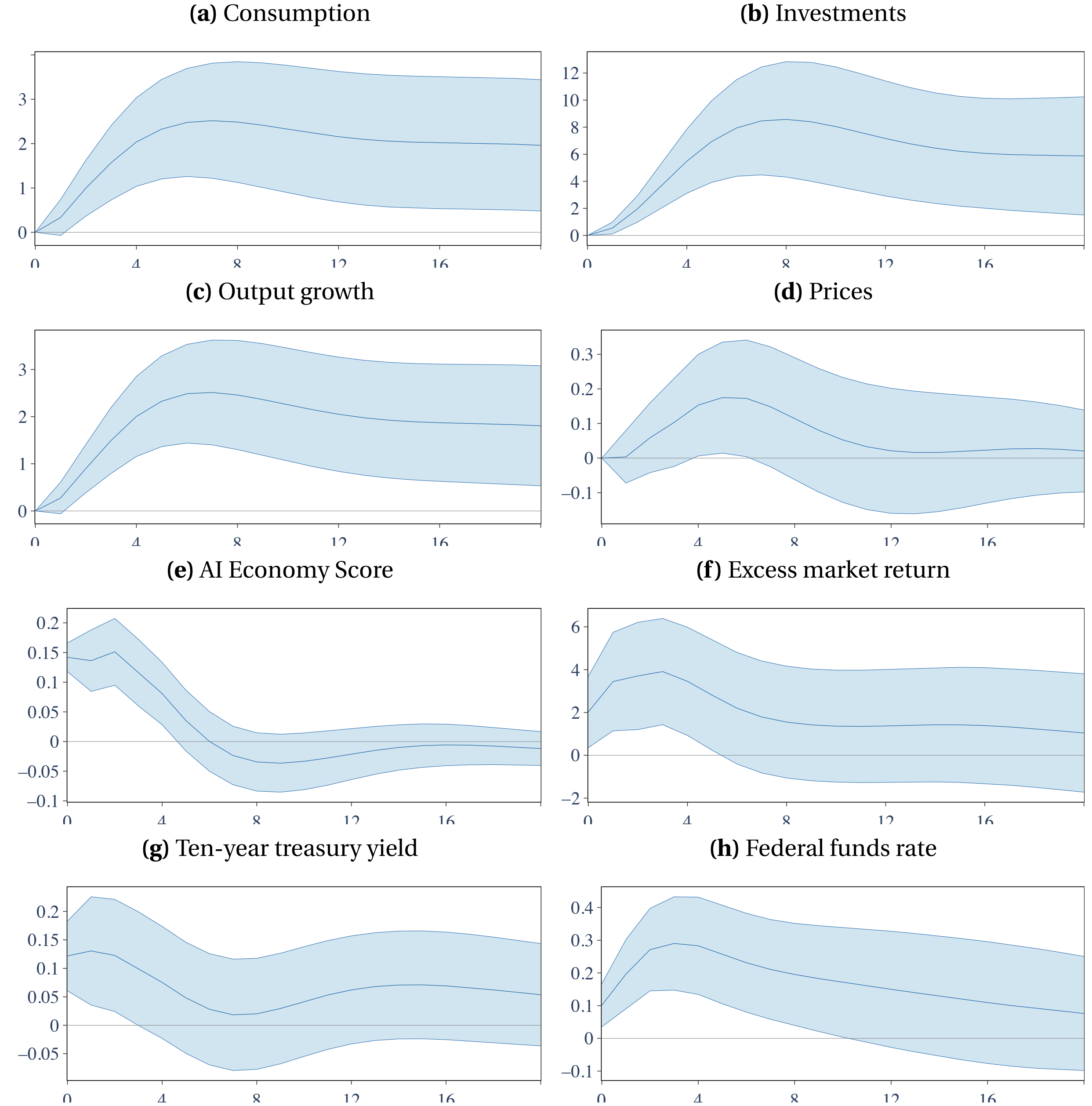

**Table 1.** Words associated with low- and high- *AI Economy Score*

This table displays the top ten 3-grams and 4-grams associated with both low and high AI Economy scores. We examine the reasons why ChatGPT assigned a significant decrease or decrease for low AI Economy scores (scores of 1-3), and similarly a significant increase or increase for high AI Economy scores (scores of 5-7); a neutral score of 4 is excluded. To ensure consistency, we lemmatize each word to account for differing grammatical noun and verb forms. Additionally, we restrict the selection of words to those found in the English dictionary, excluding stop words and proper nouns. We further exclude (i) the sentiment- and grading-vocabulary ChatGPT uses to justify a score (e.g., "optimistic," "pessimistic," "near term," "management expects") and (ii) neutral numeric-magnitude phrasing ("single digit," "double digit"). To ensure each column reflects language that is genuinely characteristic of its score bucket rather than phrasing common to both, n-grams are ranked by their distinctiveness: the difference between an n-gram's frequency per 1,000 explanations in the focal bucket and its frequency per 1,000 explanations in the opposite bucket. Only n-grams appearing at least 20 times in the focal bucket are eligible.

Panel A: Top ten 3-grams

| Low AI Economy Score | High AI Economy Score |
|---|---|
| weak consumer demand | strong consumer demand |
| supply chain disruption | strong north american |
| weak consumer spending | market share gain |
| elongate sale cycle | north american demand |
| consumer discretionary spending | tight labor market |
| tariff driven uncertainty | supply chain labor |
| soft consumer demand | continue strong demand |
| ongoing supply chain | input cost inflation |
| inflation rise interest | strong customer demand |
| even though demand | robust consumer demand |

Panel B: Top ten 4-grams

| Low AI Economy Score | High AI Economy Score |
|---|---|
| inflation supply chain disruption | strong north american demand |
| persistent inflation supply chain | continue strong consumer demand |
| low natural gas price | strong broad base demand |
| even though consumer demand | low cost natural gas |
| ongoing supply chain disruption | strong consumer demand rise |
| tariff drive cost pressure | demand market share gain |
| persistent supply chain disruption | boost net interest income |
| inflation drive cost pressure | growth market share gain |
| north american light vehicle | robust consumer demand strong |
| inflation supply chain constraint | labor shortage supply chain |

**Table 2.** Summary Statistics: National Level

This table displays the descriptive statistics of AI-predicted managerial forecasts on economic indicators and the corresponding realized economic indicators at the national level. We obtain ChatGPT's responses on various aspects of economic activities from earnings call transcripts using the two-pass procedure described in Section 3.1, including forecasts on the US economy, quantity of products, number of employees, and wages. Each score is an integer from 1 to 7, with 4 as the neutral midpoint. We convert the firm-level AI-predicted scores to aggregate-level scores by calculating the equally-weighted average across firms in each quarter. The national-level realized economic indicators are obtained from Federal Reserve Economic Data (FRED). We define the realized economic indicators as the logarithm of the next period (t+1) over the last period (t-1) in our regression analyses. *Term Spread* is the difference between ten-year and three-month constant-maturity Treasury yield. *Real FFR* is the real federal funds rate and the *GZ Spread* is the credit spread index. Variable definitions are in Appendix A.

| | N | Mean | SD | p25 | Median | p75 |
|---|---|---|---|---|---|---|
| *AI Economy Score* | 80 | 4.252 | 0.427 | 4.118 | 4.331 | 4.479 |
| *AI Economy Score_Entity Neutered* | 80 | 3.994 | 0.620 | 3.762 | 4.131 | 4.354 |
| *SPF Forecasted rGDP* | 80 | 0.021 | 0.022 | 0.017 | 0.023 | 0.029 |
| *Real GDP* | 80 | 0.010 | 0.018 | 0.006 | 0.012 | 0.015 |
| *Industrial Production* | 80 | 0.001 | 0.032 | -0.006 | 0.007 | 0.015 |
| *Payroll Employment* | 80 | 0.004 | 0.021 | 0.003 | 0.007 | 0.009 |
| *Wage* | 80 | 0.014 | 0.005 | 0.009 | 0.014 | 0.017 |
| *Term Spread* | 80 | 0.012 | 0.013 | 0.002 | 0.014 | 0.022 |
| *Real FFR* | 80 | 0.017 | 0.020 | 0.001 | 0.005 | 0.038 |
| *GZ Spread* | 80 | 0.021 | 0.011 | 0.015 | 0.018 | 0.025 |

**Table 3.** AI Economy Prediction and Real GDP Growth: National Level

This table presents the coefficients obtained by regressing the growth of the Real GDP for the next quarter/future quarters on AI-predicted scores at the national level. Panel A displays the results for the next quarter. The dependent variables are defined as the logarithm of the next period $t+1$ over the previous period $t-1$. Panel B reports the results for longer horizons. Here, the dependent variables are defined as the logarithm of the upcoming $n$-th quarter (up to six quarters) $t+n$ compared to the previous quarter $t-1$. Four lags of the dependent variables are controlled for in all specifications. Variable definitions are in Appendix A.

Panel A: Next Quarter

| | (1) | (2) | (3) |
|---|---|---|---|
| | *Real GDP:* Next Quarter | | |
| *Term Spread* | 0.208 | -0.0268 | 0.101 |
| | (0.48) | (-0.07) | (0.23) |
| *Real FFR* | -0.0680 | 0.0876 | 0.0501 |
| | (-0.30) | (0.39) | (0.25) |
| *GZ Spread* | -1.259*** | | -0.500 |
| | (-4.61) | | (-1.28) |
| *AI Economy Score* | | 0.0304*** | 0.0207*** |
| | | (6.86) | (3.67) |
| R-squared | 0.438 | 0.471 | 0.486 |
| Observations | 80 | 80 | 80 |

Panel B: Long horizons

| | (1) | (2) | (3) | (4) | (5) |
|---|---|---|---|---|---|
| | | | *Real GDP* | | |
| | 2 quarters | 3 quarters | 4 quarters | 5 quarters | 6 quarters |
| *Term Spread* | -0.121 | -0.193 | -0.442 | -0.759** | -1.080*** |
| | (-0.39) | (-0.50) | (-1.20) | (-2.45) | (-3.88) |
| *Real FFR* | -0.0163 | -0.178 | -0.447*** | -0.721*** | -1.080*** |
| | (-0.07) | (-0.90) | (-2.94) | (-4.40) | (-5.43) |
| *GZ Spread* | -0.194 | -0.523 | -0.728 | -0.720* | -0.975* |
| | (-0.58) | (-1.56) | (-1.53) | (-1.67) | (-1.79) |
| *AI Economy Score* | 0.0327*** | 0.0287*** | 0.0225** | 0.0211** | 0.00905 |
| | (3.21) | (3.26) | (2.08) | (2.21) | (0.72) |
| R-squared | 0.546 | 0.566 | 0.545 | 0.565 | 0.565 |
| Observations | 79 | 78 | 77 | 76 | 75 |

**Table 4.** Alternative Economic Indicators: National Level

This table presents the coefficients obtained by regressing the growth of the alternative economic predictors, *Industrial Production*, *Employment*, and *Wages*, for the next quarter on *AI Economy Score* at the national level. Panel A displays the results for predictions for the next quarter, while Panel B shows the results for long-term predictions. The dependent variables are defined as the logarithm of the upcoming $n$-th quarter (up to ten quarters) $t+n$ over the previous quarter $t-1$. Four lags of the dependent variables are controlled for in all specifications. Variable definitions are in Appendix A.

Panel A: Next Quarter

| | (1) | (2) | (3) | (4) | (5) | (6) | (7) | (8) | (9) |
|---|---|---|---|---|---|---|---|---|---|
| | *Industrial Production* | | | *Employment* | | | *Wages* | | |
| | Forecasting Period: Next Quarter | | | Forecasting Period: Next Quarter | | | Forecasting Period: Next Quarter | | |
| *Term Spread* | 1.826*** | 1.190** | 1.738** | 0.514 | 0.332 | 0.385 | -0.00735 | 0.00242 | -0.0134 |
| | (2.83) | (2.06) | (2.62) | (0.81) | (0.61) | (0.64) | (-0.13) | (0.05) | (-0.30) |
| *Real FFR* | 0.439 | 0.722** | 0.601* | 0.0495 | 0.199 | 0.183 | -0.0375 | -0.0139 | -0.00802 |
| | (1.27) | (2.05) | (1.99) | (0.15) | (0.59) | (0.55) | (-1.04) | (-0.50) | (-0.31) |
| *GZ Spread* | -2.909*** | | -1.994*** | -1.175*** | | -0.217 | -0.155*** | | 0.0861 |
| | (-6.13) | | (-3.42) | (-3.59) | | (-0.32) | (-4.58) | | (1.64) |
| *AI Economy Score* | | 0.0624*** | 0.0254** | | 0.0287*** | 0.0245 | | 0.00491*** | 0.00664*** |
| | | (5.53) | (2.19) | | (4.46) | (1.58) | | (6.26) | (4.39) |
| R-squared | 0.488 | 0.438 | 0.512 | 0.262 | 0.311 | 0.313 | 0.775 | 0.822 | 0.826 |
| Observations | 80 | 80 | 80 | 80 | 80 | 80 | 80 | 80 | 80 |

*(continued)*

Panel B: Long-term

| | (1) | (2) | (3) | (4) | (5) | (6) | (7) | (8) | (9) |
|---|---|---|---|---|---|---|---|---|---|
| | | | | | *Industrial Production* | | | | |
| | 2 quarters | 3 quarters | 4 quarters | 5 quarters | 6 quarters | 7 quarters | 8 quarters | 9 quarters | 10 quarters |
| *Term Spread* | 1.720*** | 1.661** | 1.345 | 0.943 | 0.488 | 0.192 | 0.000988 | -0.0382 | 0.0805 |
| | (3.00) | (2.24) | (1.66) | (1.15) | (0.60) | (0.23) | (0.00) | (-0.05) | (0.10) |
| *Real FFR* | 0.547 | 0.437 | 0.0584 | -0.339 | -0.914 | -1.369* | -1.808** | -2.164*** | -2.313*** |
| | (1.45) | (1.08) | (0.13) | (-0.63) | (-1.45) | (-1.99) | (-2.62) | (-3.50) | (-4.62) |
| *GZ Spread* | -1.918*** | -1.673** | -1.766* | -1.465 | -1.632 | -1.501 | -1.525 | -1.580 | -1.162 |
| | (-3.14) | (-2.09) | (-1.68) | (-1.39) | (-1.36) | (-1.31) | (-1.31) | (-1.29) | (-0.98) |
| *AI Economy Score* | 0.0387** | 0.0517** | 0.0474 | 0.0493* | 0.0324 | 0.0208 | 0.00600 | -0.00864 | -0.00583 |
| | (2.02) | (2.01) | (1.53) | (1.69) | (1.04) | (0.66) | (0.18) | (-0.25) | (-0.20) |
| R-squared | 0.462 | 0.429 | 0.383 | 0.370 | 0.364 | 0.376 | 0.423 | 0.480 | 0.538 |
| Observations | 79 | 78 | 77 | 76 | 75 | 74 | 73 | 72 | 71 |

| | (1) | (2) | (3) | (4) | (5) | (6) | (7) | (8) | (9) |
|---|---|---|---|---|---|---|---|---|---|
| | | | | | *Employment* | | | | |
| | 2 quarters | 3 quarters | 4 quarters | 5 quarters | 6 quarters | 7 quarters | 8 quarters | 9 quarters | 10 quarters |
| *Term Spread* | 0.268 | 0.296 | 0.171 | -0.0597 | -0.337 | -0.598 | -0.826* | -1.026** | -1.155*** |
| | (0.52) | (0.50) | (0.28) | (-0.11) | (-0.64) | (-1.22) | (-1.78) | (-2.30) | (-2.78) |
| *Real FFR* | 0.113 | 0.00898 | -0.244 | -0.526* | -0.902*** | -1.253*** | -1.605*** | -1.958*** | -2.177*** |
| | (0.34) | (0.03) | (-0.84) | (-1.92) | (-3.23) | (-4.25) | (-5.16) | (-6.01) | (-7.32) |
| *GZ Spread* | 0.0348 | -0.133 | -0.464 | -0.591 | -0.970 | -1.233 | -1.493* | -1.788** | -1.592** |
| | (0.06) | (-0.27) | (-0.75) | (-0.96) | (-1.24) | (-1.55) | (-1.79) | (-2.01) | (-2.01) |
| *AI Economy Score* | 0.0372** | 0.0415*** | 0.0382** | 0.0377** | 0.0285 | 0.0197 | 0.0106 | -0.00136 | 0.0000126 |
| | (2.62) | (3.26) | (2.60) | (2.62) | (1.57) | (0.97) | (0.47) | (-0.05) | (0.00) |
| R-squared | 0.360 | 0.408 | 0.419 | 0.446 | 0.462 | 0.475 | 0.506 | 0.534 | 0.564 |
| Observations | 80 | 79 | 78 | 77 | 76 | 75 | 74 | 73 | 72 |

*(continued)*

| | (1) | (2) | (3) | (4) | (5) | (6) | (7) | (8) | (9) |
|---|---|---|---|---|---|---|---|---|---|
| | | | | | *Wages* | | | | |
| | 2 quarters | 3 quarters | 4 quarters | 5 quarters | 6 quarters | 7 quarters | 8 quarters | 9 quarters | 10 quarters |
| *Term Spread* | -0.0615 | -0.161** | -0.305*** | -0.489*** | -0.715*** | -0.962*** | -1.205*** | -1.463*** | -1.701*** |
| | (-1.13) | (-2.11) | (-3.12) | (-3.78) | (-4.67) | (-5.41) | (-5.99) | (-6.71) | (-7.13) |
| *Real FFR* | -0.0406 | -0.105** | -0.187*** | -0.309*** | -0.448*** | -0.595*** | -0.746*** | -0.898*** | -1.060*** |
| | (-1.28) | (-2.02) | (-2.76) | (-3.44) | (-4.06) | (-4.71) | (-5.13) | (-5.87) | (-6.60) |
| *GZ Spread* | 0.133* | 0.183 | 0.273* | 0.256 | 0.318 | 0.317 | 0.314 | 0.338 | 0.341 |
| | (1.76) | (1.63) | (1.96) | (1.42) | (1.53) | (1.35) | (1.17) | (1.19) | (1.09) |
| *AI Economy Score* | 0.0106*** | 0.0140*** | 0.0179*** | 0.0186*** | 0.0210*** | 0.0219*** | 0.0226*** | 0.0241*** | 0.0251** |
| | (5.18) | (4.20) | (4.29) | (3.32) | (3.19) | (3.01) | (2.68) | (2.72) | (2.60) |
| R-squared | 0.847 | 0.844 | 0.848 | 0.835 | 0.831 | 0.828 | 0.830 | 0.839 | 0.843 |
| Observations | 79 | 78 | 77 | 76 | 75 | 74 | 73 | 72 | 71 |

**Table 5.** Summary Statistics: Micro Level

This table displays the descriptive statistics of AI-predicted managerial forecasts on economic indicators and the corresponding realized economic indicators at the industry and firm levels, presented in Panels A and B, respectively. We obtain ChatGPT's responses on various aspects of economic activities from earnings call transcripts using the two-pass procedure described in Section 3.1, including forecasts on the US economy, quantity of products, number of employees, and wages. Each score is an integer from 1 to 7, with 4 as the neutral midpoint. We convert the firm-level AI-predicted scores to industry-level scores by calculating the equally-weighted average within each of the 19 sectors under the NAICS classification standard (the sector *Other Services (except Public Administration)* is merged with *Public Administration (not covered in the economic census)*). At all three levels the score comes from the question about the outlook for the US economy. The *AI Economy Score* is the raw national-level score; the *AI Industry Score* and *AI Firm Score* are the adjusted components defined in equation (2), net of the national and sector averages respectively. In Panel B, the annual statistics are reported separately for the employment sample and the wage sample, since the AI-score components are recomputed on each sample and the two samples differ substantially in size. The industry-level realized economic indicators are obtained from the Bureau of Economic Analysis (BEA), and the firm-level realized economic indicators are obtained from WRDS. We define the realized economic indicators as the logarithm of the next period (t+1) over the last period (t-1) in our regression analyses. Variable definitions are in Appendix A.

| | N | Mean | SD | p25 | Median | p75 |
|---|---|---|---|---|---|---|
| **Panel A: Industry-Level** | | | | | | |
| ***Quarterly*** | | | | | | |
| *AI Economy Score* | 1,313 | 4.260 | 0.427 | 4.181 | 4.334 | 4.484 |
| *AI Industry Score* | 1,313 | 0.087 | 0.457 | -0.141 | 0.064 | 0.308 |
| *Ind. Real GDP* | 1,313 | 0.010 | 0.048 | -0.002 | 0.013 | 0.026 |
| ***Annual*** | | | | | | |
| *AI Economy Score* | 315 | 4.240 | 0.393 | 4.116 | 4.266 | 4.468 |
| *AI Industry Score* | 315 | 0.042 | 0.415 | -0.155 | 0.055 | 0.258 |
| *Ind. Employment* | 315 | 0.016 | 0.072 | -0.007 | 0.027 | 0.054 |
| *Ind. Wage* | 315 | 0.081 | 0.084 | 0.056 | 0.089 | 0.125 |
| | | | | | | |
| **Panel B: Firm-Level Sample** | | | | | | |
| ***Quarterly*** | | | | | | |
| *AI Economy Score* | 89,774 | 4.322 | 0.361 | 4.210 | 4.354 | 4.487 |
| *AI Industry Score* | 89,774 | 0.003 | 0.238 | -0.104 | -0.032 | 0.084 |
| *AI Firm Score* | 89,774 | 0.025 | 1.064 | -0.825 | 0.237 | 0.784 |
| *Value-Add* | 89,774 | 0.028 | 0.374 | -0.078 | 0.032 | 0.143 |
| *Sales* | 89,774 | 0.032 | 0.256 | -0.052 | 0.031 | 0.117 |
| ***Annually (Employment Sample)*** | | | | | | |
| *AI Economy Score* | 30,607 | 4.285 | 0.369 | 4.186 | 4.314 | 4.512 |
| *AI Industry Score* | 30,607 | -0.000 | 0.231 | -0.083 | -0.013 | 0.063 |
| *AI Firm Score* | 30,607 | -0.014 | 0.803 | -0.514 | -0.048 | 0.547 |
| *Firm Employment* | 30,607 | 0.055 | 0.389 | -0.065 | 0.045 | 0.184 |
| ***Annually (Wage Sample)*** | | | | | | |
| *AI Economy Score* | 3,112 | 4.315 | 0.356 | 4.186 | 4.314 | 4.512 |
| *AI Industry Score* | 3,112 | 0.029 | 0.278 | -0.136 | 0.019 | 0.197 |
| *AI Firm Score* | 3,112 | -0.040 | 0.775 | -0.540 | -0.029 | 0.515 |
| *Firm Wage* | 3,112 | 0.121 | 0.294 | 0.000 | 0.112 | 0.239 |

**Table 6.** AI Economy Prediction and Real GDP Growth: Industry Level

This table reports the coefficients from regressing the growth of Real GDP for the next quarter/year on AI-predicted economy score at the industry level. We examine the predictive ability of the national level *AI Economy Score* and the *AI Industry Score*. The *AI Economy Score* is the equally-weighted average of the firm-level scores across all firms in a given period. The *AI Industry Score* is an adjusted score: it is the equally-weighted average of the firm-level scores within each of the 19 NAICS sectors, minus the national average in the same period. Because the industry score enters as a deviation from the national average, the two AI-based regressors are orthogonal by construction and their coefficients are separately identified. Panel A presents the results for the next quarter; Panel B presents results for longer horizons; The dependent variables are defined as the logarithm of the next period $t+1$ over the last period $t-1$. Four lags of dependent variables are controlled for in all specifications. Variable definitions are in Appendix A.

Panel A: Next Period

| | (1) | (2) | (3) | (4) |
|---|---|---|---|---|
| | *Ind. Real GDP*: Next Quarter | | | |
| *Term Spread* | 0.818*** | 0.450 | -0.173 | 0.744** |
| | (2.63) | (1.61) | (-0.64) | (2.35) |
| *Real FFR* | 0.0298 | 0.249 | -0.205 | 0.175 |
| | (0.17) | (1.40) | (-1.21) | (1.00) |
| *GZ Spread* | -1.762*** | | | -0.974** |
| | (-8.17) | | | (-2.39) |
| *AI Economy Score* | | 0.0391*** | | 0.0199** |
| | | (8.68) | | (2.28) |
| *AI Industry Score* | | | 0.0122*** | 0.0116*** |
| | | | (3.93) | (4.09) |
| Industry FE | Yes | Yes | Yes | Yes |
| R-squared | 0.149 | 0.148 | 0.057 | 0.165 |
| Observations | 1313 | 1313 | 1313 | 1313 |

Panel B: Long Horizons

| | (1) | (2) | (3) | (4) | (5) | (6) |
|---|---|---|---|---|---|---|
| | *Ind. Real GDP*: Long Horizons | | | | | |
| | 2 quarters | 3 quarters | 4 quarters | 8 quarters | 12 quarters | 16 quarters |
| *Term Spread* | 0.484* | 0.457 | 0.102 | -1.503*** | -1.144*** | -0.0784 |
| | (1.73) | (1.38) | (0.31) | (-4.79) | (-3.58) | (-0.22) |
| *Real FFR* | 0.0376 | -0.185 | -0.697*** | -2.718*** | -2.995*** | -2.009*** |
| | (0.20) | (-0.96) | (-3.99) | (-12.02) | (-14.37) | (-8.37) |
| *GZ Spread* | -0.574 | -0.894** | -1.491*** | -2.127*** | -0.956* | -0.682 |
| | (-1.38) | (-2.15) | (-3.37) | (-4.19) | (-1.87) | (-1.20) |
| *AI Economy Score* | 0.0363*** | 0.0361*** | 0.0219** | -0.0184 | 0.00410 | 0.0275** |
| | (3.73) | (3.63) | (2.10) | (-1.49) | (0.37) | (2.12) |
| *AI Industry Score* | 0.0135*** | 0.0158*** | 0.0164*** | 0.0140*** | 0.0206*** | 0.0209*** |
| | (3.37) | (3.78) | (3.70) | (3.05) | (4.72) | (4.28) |
| Industry FE | Yes | Yes | Yes | Yes | Yes | Yes |
| R-squared | 0.182 | 0.204 | 0.222 | 0.323 | 0.400 | 0.426 |
| Observations | 1295 | 1277 | 1259 | 1187 | 1115 | 1043 |

**Table 7.** Alternative Economic Indicators: Industry Level

This table reports the coefficients from regressing the growth of alternative economic predictors (excluding the real GDP) for the next year on AI-predicted scores at the industry level. We examine the predictive ability of the national level *AI Economy Score* and the *AI Industry Score*. The *AI Economy Score* is the equally-weighted average of the firm-level scores across all firms in a given period. The *AI Industry Score* is an adjusted score: it is the equally-weighted average of the firm-level scores within each of the 19 NAICS sectors, minus the national average in the same period. The two AI-based regressors are orthogonal by construction. Panel A presents the results on employment; Panel B presents the results on wages. The dependent variables are defined as the logarithm of the next period $t+1$ over the last period $t-1$. Two lags of dependent variables are controlled for in all specifications. Variable definitions are in Appendix A.

Panel A: Employment

| | (1) | (2) | (3) | (4) |
|---|---|---|---|---|
| | *Ind. Employment:* Next Year | | | |
| *Term Spread* | 2.785*** | 1.448** | -0.427 | 2.521*** |
| | (3.61) | (2.05) | (-0.62) | (3.41) |
| *Real FFR* | 0.833** | 0.943** | -0.297 | 1.028** |
| | (2.03) | (2.28) | (-0.69) | (2.44) |
| *GZ Spread* | -4.843*** | | | -2.631** |
| | (-8.97) | | | (-2.17) |
| *AI Economy Score* | | 0.0939*** | | 0.0417 |
| | | (8.79) | | (1.65) |
| *AI Industry Score* | | | 0.0396*** | 0.0279** |
| | | | (2.86) | (2.28) |
| Industry FE | Yes | Yes | Yes | Yes |
| R-squared | 0.311 | 0.310 | 0.158 | 0.336 |
| Observations | 315 | 315 | 315 | 315 |

Panel B: Wages

| | (1) | (2) | (3) | (4) |
|---|---|---|---|---|
| | *Ind. Wages:* Next Year | | | |
| *Term Spread* | 0.957 | -0.675 | -2.848*** | 0.926 |
| | (1.02) | (-0.80) | (-3.49) | (1.00) |
| *Real FFR* | 0.149 | 0.262 | -1.105** | 0.372 |
| | (0.30) | (0.52) | (-2.20) | (0.72) |
| *GZ Spread* | -5.693*** | | | -3.882*** |
| | (-9.00) | | | (-2.79) |
| *AI Economy Score* | | 0.107*** | | 0.0322 |
| | | (8.56) | | (1.11) |
| *AI Industry Score* | | | 0.0518*** | 0.0397*** |
| | | | (3.51) | (3.16) |
| Industry FE | Yes | Yes | Yes | Yes |
| R-squared | 0.365 | 0.357 | 0.224 | 0.396 |
| Observations | 315 | 315 | 315 | 315 |

**Table 8.** AI Economy Prediction: Firm Level

This table reports the coefficients from regressing the growth of the economic predictors for the next period on AI-predicted scores at the firm level. All three AI-based regressors are built from the same underlying variable, the score extracted from the question about the outlook for the US economy, and enter as the orthogonal components defined in equation (2). The *AI Economy Score* is the average of that score across all firms in a given quarter. The *AI Industry Score* is an adjusted score, equal to the equally-weighted NAICS sector average minus the national average. The *AI Firm Score* is also an adjusted score, equal to the individual firm's score minus the average across firms in the same sector. The three regressors are uncorrelated by construction, so their coefficients are separately identified. Panel A presents the results on value-added; Panel B presents results on sales; Panel C presents results on employment; Panel D presents results on wages. The dependent variables are defined as the logarithm of the next period $t+1$ over the last period $t-1$. Control variables include *Size*, *Book-to-Market*, and *Tangibility*. Two lags of dependent variables are controlled for in all specifications Variable definitions are in Appendix A..

Panel A: Value-Add

| | (1) | (2) | (3) | (4) | (5) |
|---|---|---|---|---|---|
| | *Value-Add:* Next quarter | | | | |
| *GZ Spread* | -4.573*** | | | | -0.957 |
| | (-5.56) | | | | (-0.68) |
| *AI Economy Score* | | 0.104*** | | | 0.0857** |
| | | (5.47) | | | (2.20) |
| *AI Industry Score* | | | 0.0802*** | | 0.0772*** |
| | | | (7.25) | | (6.80) |
| *AI Firm Score* | | | | 0.0274*** | 0.0268*** |
| | | | | (14.89) | (15.15) |
| Firm FE | Yes | Yes | Yes | Yes | Yes |
| R-squared | 0.351 | 0.353 | 0.347 | 0.350 | 0.360 |
| Observations | 89,774 | 89,774 | 89,774 | 89,774 | 89,774 |

Panel B: Sales

| | (1) | (2) | (3) | (4) | (5) |
|---|---|---|---|---|---|
| | *Sales:* Next quarter | | | | |
| *GZ Spread* | -3.329*** | | | | -1.188 |
| | (-4.26) | | | | (-1.01) |
| *AI Economy Score* | | 0.0721*** | | | 0.0507* |
| | | (4.52) | | | (1.70) |
| *AI Industry Score* | | | 0.0292** | | 0.0271** |
| | | | (2.24) | | (2.11) |
| *AI Firm Score* | | | | 0.0184*** | 0.0179*** |
| | | | | (14.81) | (15.30) |
| Firm FE | Yes | Yes | Yes | Yes | Yes |
| R-squared | 0.330 | 0.331 | 0.324 | 0.328 | 0.337 |
| Observations | 89,774 | 89,774 | 89,774 | 89,774 | 89,774 |

*(continued)*

Panel C: Employment

| | (1) | (2) | (3) | (4) | (5) |
|---|---|---|---|---|---|
| | | | *Firm Employment:* Next Year | | |
| *GZ Spread* | -2.808*** | | | | 0.885 |
| | (-5.09) | | | | (0.61) |
| *AI Economy Score* | | 0.0603*** | | | 0.0780** |
| | | (5.05) | | | (2.58) |
| *AI Industry Score* | | | 0.0358** | | 0.0494*** |
| | | | (2.77) | | (3.49) |
| *AI Firm Score* | | | | 0.0401*** | 0.0425*** |
| | | | | (7.77) | (9.60) |
| Firm FE | Yes | Yes | Yes | Yes | Yes |
| R-squared | 0.336 | 0.337 | 0.335 | 0.340 | 0.343 |
| Observations | 30,607 | 30,607 | 30,607 | 30,607 | 30,607 |

Panel D: Wage

| | (1) | (2) | (3) | (4) | (5) |
|---|---|---|---|---|---|
| | | | *Firm Wages:* Next Year | | |
| *GZ Spread* | -2.048* | | | | -0.261 |
| | (-1.96) | | | | (-0.08) |
| *AI Economy Score* | | 0.0436 | | | 0.0172 |
| | | (1.62) | | | (0.21) |
| *AI Industry Score* | | | 0.0660*** | | 0.0683*** |
| | | | (3.38) | | (3.26) |
| *AI Firm Score* | | | | 0.0338*** | 0.0364*** |
| | | | | (4.55) | (4.74) |
| Firm FE | Yes | Yes | Yes | Yes | Yes |
| R-squared | 0.432 | 0.432 | 0.433 | 0.436 | 0.440 |
| Observations | 3,112 | 3,112 | 3,112 | 3,112 | 3,112 |

**Table 9.** AI-Predicted Economic Forecasts: Heterogeneity in Managerial Behavior Types

This table shows the coefficients obtained by regressing real GDP for the next quarter on the AI-predicted score, aggregated by manager group: Accurate Managers, Inaccurate Managers, Optimists, Pessimists, and Realists. A manager is classified as an Optimist (Pessimist) if the average bias of their past eight-quarter forecasts ranks in the top (bottom) quartile, where bias is calculated as the difference between the tercile ranking of the manager's AI Economy Score and the tercile ranking of realized real GDP. The remaining managers are classified as Realists. A manager is classified as Accurate (Inaccurate) if their forecast accuracy ranks in the top (bottom) quartile. Forecast accuracy is the R-squared obtained from regressing real GDP on the firm-level AI-predicted score using the past eight quarters of data. The dependent variable is defined as the logarithm of the next period's GDP $(t+1)$ over the previous period's GDP $(t-1)$. Variable definitions are in Appendix A.

| | (1) | (2) | (3) | (4) | (5) | (6) | (7) | (8) | (9) |
|---|---|---|---|---|---|---|---|---|---|
| | | | | | *Real GDP:* Next Quarter | | | | |
| *Term Spread* | 0.191 | 0.114 | 0.0954 | 0.186 | 0.112 | 0.163 | 0.129 | 0.0335 | 0.0350 |
| | (0.43) | (0.26) | (0.21) | (0.42) | (0.25) | (0.37) | (0.29) | (0.07) | (0.08) |
| *Real FFR* | -0.0950 | 0.0911 | 0.0155 | 0.0308 | 0.0391 | -0.0485 | 0.108 | 0.0557 | 0.119 |
| | (-0.40) | (0.46) | (0.08) | (0.16) | (0.22) | (-0.22) | (0.55) | (0.28) | (0.59) |
| *GZ Spread* | -1.260*** | -0.431 | -0.742* | -0.745 | -0.657 | -0.922** | -0.463 | -0.448 | -0.338 |
| | (-4.58) | (-1.04) | (-1.83) | (-1.38) | (-1.23) | (-2.55) | (-1.12) | (-1.06) | (-0.76) |
| *AI Economy Score* | | 0.0226*** | | | | | | | |
| | | (3.56) | | | | | | | |
| *AI Economy Score: Accurate Managers* | | | 0.0140** | | 0.0113** | | | | |
| | | | (2.65) | | (2.62) | | | | |
| *AI Economy Score: Inaccurate Managers* | | | | 0.0135 | 0.00480 | | | | |
| | | | | (1.50) | (0.52) | | | | |
| *AI Economy Score: Optimist* | | | | | | 0.00973** | | | -0.00725* |
| | | | | | | (2.01) | | | (-1.78) |
| *AI Economy Score: Pessimist* | | | | | | | 0.0213*** | | 0.0117 |
| | | | | | | | (3.43) | | (1.44) |
| *AI Economy Score: Realist* | | | | | | | | 0.0237*** | 0.0214** |
| | | | | | | | | (3.59) | (2.06) |
| R-squared | 0.440 | 0.487 | 0.469 | 0.460 | 0.471 | 0.457 | 0.488 | 0.489 | 0.498 |
| Observations | 75 | 75 | 75 | 75 | 75 | 75 | 75 | 75 | 75 |

**Table 10.** AI-Predicted Economic Forecasts: Focused vs. Diversified Firms

This table presents the AI economy prediction and the growth of realized GDP across firms with different levels of operational diversification. *AI Economy Score* is a firm-specific score derived from a single earnings call. *Business Segment HHI* is the sum of the squared ratio of a business segment's sales to total sales. *High Business Segment HHI* is an indicator variable that represents whether the *Business Segment HHI* is above the median in a quarter across all firms. *GEO HHI* is the sum of the squared ratio of the number of times a state is mentioned in a 10-K filing to the number of different states mentioned. *High GEO HHI* is an indicator variable that represents whether the *GEO HHI* is above the median in a quarter across all firms. Four lags of the dependent variables are controlled for in all specifications. Firm fixed effects are included in all specifications. Standard errors are clustered at the firm and year-quarter level. Variable definitions are in Appendix A.

| | (1) | (2) | (3) | (4) |
|---|---|---|---|---|
| | | *Real GDP:* Next Quarter | | |
| *AI Economy Score* × *Business Segment HHI* | 0.00101** | | | |
| | (2.09) | | | |
| *AI Economy Score* × *High Business Segment HHI* | | 0.000539** | | |
| | | (2.13) | | |
| *AI Economy Score* × *Geo HHI* | | | 0.000580 | |
| | | | (1.07) | |
| *AI Economy Score* × *High Geo HHI* | | | | 0.000126 |
| | | | | (0.84) |
| *Term Spread* | 0.00891 | 0.00894 | 0.00897 | 0.00895 |
| | (0.94) | (0.94) | (0.95) | (0.94) |
| *Real FFR* | 0.00173 | 0.00169 | 0.00170 | 0.00168 |
| | (0.31) | (0.30) | (0.30) | (0.30) |
| *GZ Spread* | -0.0238*** | -0.0240*** | -0.0241*** | -0.0241*** |
| | (-4.23) | (-4.28) | (-4.27) | (-4.29) |
| Firm FE | Yes | Yes | Yes | Yes |
| R-squared | 0.274 | 0.274 | 0.274 | 0.274 |
| Observations | 79,812 | 79,812 | 79,812 | 79,812 |

**Table 11.** AI Weighted Score and Real GDP Growth

This table presents the coefficients from regressions of the growth of real GDP for the next quarter and future quarters on the weighted AI-predicted score at the national level. The weighted AI-predicted score, *AI Weighted Score*, is a linear combination of 14 national-level AI scores corresponding to the 14 questions posed to the model. The weight assigned to each score in quarter $t$ is the beta loading from a firm-quarter level regression of firm sales on the 14 lagged firm-level AI scores, estimated without an intercept via ridge regression with a negligible penalty ($\lambda = 0.001$) using all observations available prior to quarter $t$, so that the weights are known at quarter $t$ without look-ahead bias. The t-statistics are based on Newey and West (1987) standard errors with two lags. All specifications control for four lags of the dependent variable. Variable definitions are in Appendix A.

| | (1) | (2) | (3) | (4) | (5) | (6) | (7) | (8) |
|---|---|---|---|---|---|---|---|---|
| | | | | *Real GDP* | | | | |
| | 1 qtr | 2 qtrs | 3 qtrs | 4 qtrs | 5 qtrs | 6 qtrs | 7 qtrs | 8 qtrs |
| *Term Spread* | 0.1297 | -0.0198 | -0.0951 | -0.3692 | -0.7087** | -1.0975*** | -1.3801*** | -1.4985*** |
| | (0.31) | (-0.06) | (-0.23) | (-1.06) | (-2.37) | (-4.15) | (-5.52) | (-5.88) |
| *Real FFR* | -0.1199 | -0.2466 | -0.3714* | -0.6021*** | -0.8779*** | -1.1751*** | -1.4193*** | -1.6067*** |
| | (-0.56) | (-1.27) | (-1.80) | (-3.26) | (-4.82) | (-6.00) | (-6.89) | (-7.52) |
| *GZ Spread* | -1.1461*** | -1.2922*** | -1.4989*** | -1.4901*** | -1.4073*** | -1.2157*** | -0.9905*** | -0.9804*** |
| | (-4.48) | (-7.76) | (-6.27) | (-5.15) | (-5.10) | (-4.60) | (-4.40) | (-4.53) |
| *AI Weighted Score* | 0.1105** | 0.1040** | 0.0933*** | 0.0761*** | 0.0953*** | 0.0889** | 0.1185** | 0.0973* |
| | (2.26) | (1.96) | (3.18) | (2.73) | (3.01) | (2.08) | (2.41) | (1.90) |
| R-squared | 0.4652 | 0.4729 | 0.5219 | 0.5208 | 0.5510 | 0.5691 | 0.5815 | 0.6204 |
| Observations | 80 | 79 | 78 | 77 | 76 | 75 | 74 | 73 |

**Table 12.** AI Economic Forecast vs. Survey Forecast: A Horse Race

This table presents the coefficients from regressing the next quarter's Real GDP growth on the AI-predicted economy score and the Survey-Forecasted economic growth at the national level. The survey-based measure, *SPF-Forecasted rGDP*, is derived from the Survey of Professional Forecasters. It uses the median forecasts of survey respondents for Real GDP in the upcoming quarter $(t+1)$ and the most recent realized RGDP for the previous quarter $(t-1)$ to construct a forecasted GDP growth. The dependent variables are defined as the logarithm of the next period $t+1$ over the last period $t-1$. Four lags of dependent variables are controlled for in all specifications. Variable definitions are in Appendix A.

| | (1) | (2) | (3) |
|---|---|---|---|
| | *Real GDP:* Next Quarter | | |
| *Term Spread* | 0.101 | 0.0648 | 0.0563 |
| | (0.23) | (0.14) | (0.12) |
| *Real FFR* | 0.0501 | -0.0788 | -0.0124 |
| | (0.25) | (-0.36) | (-0.06) |
| *GZ Spread* | -0.500 | -0.974*** | -0.668* |
| | (-1.28) | (-3.03) | (-1.92) |
| *AI Economy Score* | 0.0207*** | | 0.0110* |
| | (3.67) | | (1.98) |
| *SPF-Forecasted rGDP* | | 0.376*** | 0.249* |
| | | (2.94) | (1.90) |
| R-squared | 0.486 | 0.491 | 0.499 |
| Observations | 80 | 80 | 80 |

**Table 13.** AI Economy Forecast: Entity-Neutering tests

This table presents the coefficients from regressing next quarter's real GDP growth on the AI-predicted economy score using an entity-neutered sample. *AI Economy Score_Entity Neutered* is obtained from firms' earnings call transcripts after applying the entity-neutering procedure of Engelberg et al. (2025) to remove identifying information. The dependent variables are defined as the logarithm of the next period $t+1$ over the last period $t-1$. Four lags of dependent variables are controlled for in all specifications. Variable definitions are in Appendix A.

Panel A: Next Quarter

| | (1) | (2) | (3) |
|---|---|---|---|
| | | *Real GDP:* Next Quarter | |
| *Term Spread* | 0.208 | -0.118 | 0.126 |
| | (0.48) | (-0.33) | (0.29) |
| *Real FFR* | -0.0680 | 0.0704 | 0.0221 |
| | (-0.30) | (0.31) | (0.11) |
| *GZ Spread* | -1.259*** | | -0.774** |
| | (-4.61) | | (-2.27) |
| *AI Economy Score_Entity Neutered* | | 0.0196*** | 0.00995*** |
| | | (6.11) | (3.23) |
| R-squared | 0.438 | 0.420 | 0.470 |
| Observations | 80 | 80 | 80 |

Panel B: Long horizons

| | (1) | (2) | (3) | (4) | (5) |
|---|---|---|---|---|---|
| | | | *Real GDP* | | |
| | 2 quarters | 3 quarters | 4 quarters | 5 quarters | 6 quarters |
| *Term Spread* | -0.0792 | -0.156 | -0.405 | -0.737** | -1.069*** |
| | (-0.24) | (-0.38) | (-1.06) | (-2.35) | (-3.86) |
| *Real FFR* | -0.0617 | -0.219 | -0.491*** | -0.750*** | -1.093*** |
| | (-0.29) | (-1.13) | (-3.26) | (-4.52) | (-5.53) |
| *GZ Spread* | -0.637*** | -0.914*** | -1.090** | -0.990** | -1.097** |
| | (-2.92) | (-3.04) | (-2.44) | (-2.53) | (-2.51) |
| *AI Economy Score_Entity Neutered* | 0.0155*** | 0.0136** | 0.00951 | 0.0103* | 0.00431 |
| | (3.09) | (2.65) | (1.36) | (1.73) | (0.64) |
| R-squared | 0.513 | 0.545 | 0.529 | 0.557 | 0.564 |
| Observations | 79 | 78 | 77 | 76 | 75 |

**Table 14.** AI Economy Forecast: Out-of-Sample Test

This table reports the coefficients from regressing the growth of firm Sales (Panel A) and industry Real GDP (Panel B) for the next quarter on the current quarter's AI-based scores, using data from 2024Q2 to 2025Q4, a period that falls outside the GPT-5 mini model's training window. The AI-based scores enter as the orthogonal components defined in equation (2), all built from the question about the outlook for the US economy: the *AI Economy Score* is the national average, the *AI Industry Score* is the sector average net of the national average, and the *AI Firm Score* is the individual firm's score net of its sector average. The dependent variables are defined as the logarithm of the next period $t+1$ over the last period $t-1$. Control variables include *Term Spread*, *Real FFR*, *GZ Spread*, *Size*, *Book-to-Market*, and *Tangibility*. Four lags of dependent variables are controlled for in all specifications. Variable definitions are in Appendix A.

Panel A: Sales

| | (1) | (2) | (3) | (4) | (5) |
|---|---|---|---|---|---|
| | | | *Sales: Next quarter* | | |
| *GZ Spread* | -5.106 | | | | -1.857 |
| | (-0.59) | | | | (-0.15) |
| *AI Economy Score* | | -0.0484 | | | -0.0416 |
| | | (-1.11) | | | (-0.55) |
| *AI Industry Score* | | | -0.0000444 | | -0.00254 |
| | | | (-0.00) | | (-0.17) |
| *AI Firm Score* | | | | 0.0102** | 0.0103** |
| | | | | (3.06) | (3.10) |
| Firm FE | Yes | Yes | Yes | Yes | Yes |
| R-squared | 0.504 | 0.504 | 0.504 | 0.504 | 0.504 |
| Observations | 13,501 | 13,501 | 13,501 | 13,501 | 13,501 |

Panel B: Industry Real GDP

| | (1) | (2) | (3) | (4) |
|---|---|---|---|---|
| | | *Ind. Real GDP: Next quarter* | | |
| *Term Spread* | 0.00559 | -0.122 | -0.287 | 1.351 |
| | (0.00) | (-0.10) | (-0.26) | (0.82) |
| *Real FFR* | 0.755 | 0.707 | 0.516 | 2.090 |
| | (0.56) | (0.47) | (0.39) | (1.16) |
| *GZ Spread* | 1.834 | | | 5.578 |
| | (0.83) | | | (1.63) |
| *AI Economy Score* | | -0.00227 | | -0.0368 |
| | | (-0.14) | | (-1.51) |
| *AI Industry Score* | | | 0.0158** | 0.0151** |
| | | | (2.08) | (2.00) |
| Industry FE | Yes | Yes | Yes | Yes |
| R-squared | 0.536 | 0.532 | 0.553 | 0.565 |
| Observations | 108 | 108 | 108 | 108 |

**Table 15.** Comparison with Alternative Generative AI Model

This table presents the coefficients obtained by regressing the growth of the Real GDP for the next quarter/future quarters on the AI-predicted score obtained from by the Gemma model at the aggregate level. Panel A presents the results for one-quarter-ahead forecasts. The dependent variables are defined as the logarithm of the next period $t+1$ over the last period $t-1$. Panel B reports the corresponding results for longer-horizon forecasts. Here, the dependent variables are defined as the logarithm of the upcoming n-th quarter (up to six quarters) $t+n$ compared to the previous quarter $t-1$. Four lags of dependent variables are controlled for in all specifications. Variable definitions are in Appendix A.

Panel A: Gemma Model for the Next Quarter

| | (1) | (2) | (3) |
|---|---|---|---|
| | *Real GDP:* Next Quarter | | |
| *Term Spread* | 0.208 | -0.126 | 0.113 |
| | (0.48) | (-0.35) | (0.26) |
| *Real FFR* | -0.0680 | 0.0786 | 0.0261 |
| | (-0.30) | (0.34) | (0.13) |
| *GZ Spread* | -1.259*** | | -0.753** |
| | (-4.61) | | (-2.15) |
| *AI Economy Score_Gemma* | | 0.0199*** | 0.0103*** |
| | | (6.26) | (3.19) |
| R-squared | 0.438 | 0.425 | 0.469 |
| Observations | 80 | 80 | 80 |

Panel B: Gemma Model for Long Horizons

| | (1) | (2) | (3) | (4) | (5) |
|---|---|---|---|---|---|
| | | | *Real GDP* | | |
| | 2 quarters | 3 quarters | 4 quarters | 5 quarters | 6 quarters |
| *Term Spread* | -0.0989 | -0.165 | -0.397 | -0.724** | -1.041*** |
| | (-0.30) | (-0.40) | (-1.00) | (-2.22) | (-3.63) |
| *Real FFR* | -0.0556 | -0.221 | -0.503*** | -0.764*** | -1.114*** |
| | (-0.26) | (-1.15) | (-3.34) | (-4.51) | (-5.48) |
| *GZ Spread* | -0.603** | -0.927*** | -1.168** | -1.090** | -1.243** |
| | (-2.59) | (-2.87) | (-2.37) | (-2.58) | (-2.62) |
| *AI Economy Score_Gemma* | 0.0160*** | 0.0131** | 0.00787 | 0.00821 | 0.00143 |
| | (3.05) | (2.44) | (1.01) | (1.28) | (0.19) |
| R-squared | 0.513 | 0.540 | 0.523 | 0.550 | 0.561 |
| Observations | 79 | 78 | 77 | 76 | 75 |

**Table 16.** Controlling for Loan Spread

This table presents the coefficients obtained by regressing the growth of the Real GDP for the next quarter/future quarters on the AI-predicted score obtained from the GPT-5 mini model at the aggregate level. Compared to the main setting in Table 3, this setting controls for *Loan Spread* proposed by Saunders et al. (2025) in quarter $t$. Panel A displays the results for the next quarter. The dependent variables are defined as the logarithm of the next period $t+1$ over the last period $t-1$. Panel B reports the results for longer horizons. Here, the dependent variables are defined as the logarithm of the upcoming n-th quarter (up to six quarters) $t+n$ compared to the previous quarter $t1$. Four lags of dependent variables are controlled for in all specifications. Variable definitions are in Appendix A.

Panel A: Next Quarter

| | (1) | (2) | (3) |
|---|---|---|---|
| | *Real GDP:* Next Quarter | | |
| *Term Spread* | 0.358 | 0.269 | 0.297 |
| | (0.69) | (0.56) | (0.58) |
| *Real FFR* | -0.0562 | 0.0450 | 0.0587 |
| | (-0.29) | (0.26) | (0.32) |
| *Loan Spread* | -0.103 | -0.136 | -0.0927 |
| | (-0.54) | (-0.83) | (-0.62) |
| *GZ Spread* | -1.205*** | | -0.217 |
| | (-3.29) | | (-0.77) |
| *AI Economy Score* | | 0.0298*** | 0.0267*** |
| | | (5.14) | (4.37) |
| R-squared | 0.464 | 0.523 | 0.525 |
| Observations | 66 | 66 | 66 |

Panel B: Long horizons

| | (1) | (2) | (3) | (4) | (5) |
|---|---|---|---|---|---|
| | | | *Real GDP* | | |
| | 2 quarters | 3 quarters | 4 quarters | 5 quarters | 6 quarters |
| *Term Spread* | 0.117 | 0.0890 | -0.134 | -0.408 | -0.731*** |
| | (0.31) | (0.20) | (-0.33) | (-1.31) | (-3.18) |
| *Real FFR* | -0.0536 | -0.176 | -0.430** | -0.693*** | -1.019*** |
| | (-0.36) | (-1.06) | (-2.63) | (-3.78) | (-4.74) |
| *Loan Spread* | -0.159 | -0.0601 | 0.0278 | 0.0872 | 0.181 |
| | (-0.70) | (-0.32) | (0.16) | (0.56) | (1.51) |
| *GZ Spread* | 0.526 | 0.194 | -0.0526 | -0.0425 | -0.421 |
| | (0.65) | (0.32) | (-0.09) | (-0.08) | (-0.82) |
| *AI Economy Score* | 0.0469*** | 0.0466*** | 0.0422*** | 0.0428*** | 0.0310** |
| | (2.86) | (3.31) | (3.15) | (3.56) | (2.43) |
| R-squared | 0.619 | 0.651 | 0.635 | 0.668 | 0.667 |
| Observations | 66 | 66 | 66 | 66 | 66 |

**Table 17.** Controlling for Managerial Sentiment in Earnings Calls

This table presents the coefficients obtained by regressing the growth of the Real GDP for the next quarter/future quarters on the AI-predicted score obtained from the GPT-5 mini model at the aggregate level. Compared to the main setting in Table 3, this setting controls for conference call *Sentiment*, constructed based on the Loughran-McDonald dictionary in quarter $t$. Panel A displays the results for the next quarter. The dependent variables are defined as the logarithm of the next period $t+1$ over the last period $t-1$. Panel B reports the results for longer horizons. Here, the dependent variables are defined as the logarithm of the upcoming n-th quarter (up to six quarters) $t+n$ compared to the previous quarter $t1$. Four lags of dependent variables are controlled for in all specifications. Variable definitions are in Appendix A.

Panel A: Next Quarter

| | (1) | (2) | (3) |
|---|---|---|---|
| | | *Real GDP:* Next Quarter | |
| *Term Spread* | 0.299 | 0.0119 | 0.141 |
| | (0.64) | (0.03) | (0.34) |
| *Real FFR* | 0.00327 | 0.201 | 0.176 |
| | (0.01) | (0.78) | (0.73) |
| *Sentiment* | -0.00236 | -0.00778* | -0.00735* |
| | (-0.64) | (-1.82) | (-1.81) |
| *GZ Spread* | -0.0157*** | | -0.00512 |
| | (-3.08) | | (-1.41) |
| *AI Economy Score* | | 0.0191*** | 0.0149*** |
| | | (3.96) | (4.48) |
| R-squared | 0.466 | 0.544 | 0.557 |
| Observations | 73 | 73 | 73 |

Panel B: Long horizons

| | (1) | (2) | (3) | (4) | (5) |
|---|---|---|---|---|---|
| | | | *Real GDP* | | |
| | 2 quarters | 3 quarters | 4 quarters | 5 quarters | 6 quarters |
| *Term Spread* | -0.134 | -0.204 | -0.460 | -0.785*** | -1.100*** |
| | (-0.55) | (-0.63) | (-1.47) | (-2.95) | (-4.53) |
| *Real FFR* | 0.131 | -0.0288 | -0.324** | -0.615*** | -0.983*** |
| | (0.56) | (-0.14) | (-2.23) | (-3.46) | (-4.82) |
| *Sentiment* | -0.0112** | -0.0117*** | -0.0119*** | -0.0119*** | -0.0110*** |
| | (-2.39) | (-2.87) | (-3.42) | (-2.90) | (-3.39) |
| *GZ Spread* | -0.000225 | -0.00354 | -0.00534 | -0.00498 | -0.00789 |
| | (-0.06) | (-1.12) | (-1.28) | (-1.25) | (-1.58) |
| *AI Economy Score* | 0.0241*** | 0.0228*** | 0.0204*** | 0.0199*** | 0.0140*** |
| | (3.17) | (4.15) | (5.10) | (4.34) | (3.31) |
| R-squared | 0.657 | 0.665 | 0.635 | 0.648 | 0.634 |
| Observations | 73 | 73 | 73 | 73 | 73 |

## Appendix A: Definitions of Variables

All AI scores are constructed with the two-pass procedure described in Section 3.1: an extraction pass that reads the transcript in segments and returns the sentences relevant to each topic, followed by a scoring pass that reads only the extracted sentences and issues one request per conference call. The *AI Economy Score* comes from a dedicated pipeline for the US economy. The remaining thirteen aspects are optimism about the global economy, the firm's financial prospects, the industry's financial prospects, earnings, revenue, investments, the cost of capital, wages and salaries, number of employees, production quantity, demand for their products or services, output prices, and input prices; these are extracted and scored jointly. Every score is an integer from 1 to 7, where 1 denotes a severe expected deterioration, 4 denotes mixed or offsetting signals, and 7 denotes a strong expected improvement, together with a one-sentence explanation. All scores are therefore defined at the firm-quarter level directly. A topic with no extracted sentences, or with sentences that support no directional inference, returns a null score and is treated as missing rather than as a zero. The full prompts and the intermediate filtering steps are reported in the Internet Appendix.

| Variable | Definition |
|---|---|
| *AI Economy Score* | The equally-weighted average of the firm-level score obtained from the question related to US economy for each quarter $t$. |
| *AI Economy Score_Entity Neutered* | The *AI Economy Score* constructed on anonymized transcripts, following the two-step procedure in Section 7.2: paragraph-level paraphrasing that removes firm, industry, product, and date information, followed by regular-expression and named-entity redaction. Calls whose identity GPT-5 nano can still recover are dropped, leaving about 10% of the baseline sample. Equally-weighted average across firms for each quarter $t$. |
| *AI Economy Score_Gemma* | The equally-weighted average of the firm-level score obtained from the question related to the US economy using the Gemma model (developed by Google) for each quarter $t$. |
| *AI Firm Score* | The firm-level score for the US-economy question, net of the equally-weighted average across firms in the same NAICS sector in that quarter (see Equation (2)). |
| *AI Industry Score* | The equally-weighted average of the firm-level US-economy score across firms within a NAICS sector, net of the equally-weighted national average in that quarter (see Equation (2)). |

*(continued)*

| Variable | Definition |
|---|---|
| *AI Weighted Score* | We first compute firm-level scores for the 14 aspects. Next, we calculate the equally weighted average of these scores for each quarter $t$ to obtain the corresponding national-level scores. The *AI Weighted Score* is then derived from the national-level scores, where the weight assigned to each score in each quarter is the beta loading estimated from past firm-quarter-level observations, using firm sales as the dependent variable. |
| *AI [Variable] Score* | We first obtain the firm-level scores: *AI Firm Employment Score*, *AI Firm Production Score*, and *AI Firm Wage Score*. We then calculate the equally-weighted average of these scores for each quarter $t$ to obtain the corresponding national-level scores. |
| *Book-to-Market* | Follow the definition in Fama and French (1992). |
| *Firm EBIT* | The logarithm of firms' earnings before interest and taxes in the next period $(t+1)$ over the last period $(t-1)$. |
| *Firm Employment* | The logarithm of firms' number of employees in the next period $(t+1)$ over the last period $(t-1)$. |
| *Firm Sales* | The logarithm of firms' sales over total assets in the next period $(t+1)$ over the last period $(t-1)$. |
| *GZ Spread* | The equally-weighted GZ Spread defined by Gilchrist and Zakrajšek (2012) within each quarter $t$. |
| *Industrial Production* | The logarithm of Industrial Production Index in the next period $(t+1)$ over the last period $(t-1)$. |
| *Ind. Employment* | The logarithm of Industrial Nonfarm Payroll Employment in the next period $(t+1)$ over the last period $(t-1)$. |
| *Ind. RGDP* | The logarithm of Industrial Real GDP in the next period $(t+1)$ over the last period $(t-1)$. |
| *Ind. Wage* | The logarithm of Industrial Wages in the next period $(t+1)$ over the last period $(t-1)$. |
| *Payroll Employment* | The logarithm of Nonfarm Payroll Employment in the next period $(t+1)$ over the last period $(t-1)$. |
| *Real FFR* | The equally-weighted monthly real federal funds rate within each quarter $t$. |
| *Real GDP* | The logarithm of Real GDP in the next period $(t+1)$ over the last period $(t-1)$. |
| *Tangibility* | The Compustat item Property, Plant, and Equipment - Total (Net), scaled by total assets at the end of the quarter. |
| *Term Spread* | The equally-weighted average of the monthly difference between the ten-year and the three-month constant-maturity Treasury yields within each quarter $t$. |

*(continued)*

| **Variable** | **Definition** |
| --- | --- |
| *Sentiment* | The number of positive words minus the number of negative words, scaled by the sum of positive and negative words, in each earnings call, where words are classified using the **?** dictionary. Aggregated to the national level by averaging across firms within each quarter $t$. |
| *Size* | The natural logarithm of total book assets at the end of the quarter. |
| *SPF-Forecasted rGDP* | It is the real GDP Growth rate forecasted by the survey respondents. For the survey conducted at quarter $t$, we take the median forecasts of real GDP among all the respondents for quarter $t+1$, then construct the forecasted real GDP growth rate using the median forecast for $t+1$ minus the realized real GDP in quarter $t-1$, divided by the realized real GDP in quarter $t-1$. |
| *Unemployment* | The logarithm of Unemployment Rate in the next period $(t+1)$ over the last period $(t-1)$. |
| *Value-Add* | Difference between sales and cost of goods sold. |

Internet Appendix of "Generative AI, Managerial Expectations, and Economic Activity*"

## Internet Appendix: Examples of high and low *AI Economy Score*

| Category | Example Texts from Conference Call Transcripts |
|---|---|
| *Very optimistic (Score = 7)* | "Demand for our premium cabin has been remarkable across all entities, with premium paid load factor and RASM exceeding 2019 levels. And I'd like to underscore, we see a strong demand environment this summer, and we're highly confident that that will continue going forward. Looking ahead, we feel great about the industry and what's to come for American."<br><br>"We are more optimistic about automotive than we were – we ever were. We expect the revenues to grow very nicely, but not in the short term. As you know, automotive takes at least a couple of years, depending on the application. The interest level on our products is very high, but probably the most important, the conversion of an opportunity into a design win is a lot higher than we've ever seen in any other market."<br><br>"The large projects overall are really strong. I would even say if you look at any macro indicator, ABI, Dodge and so forth, the amount of large projects that are starting with some of the stimulus to go, whether it's new factories going up, whether chips, data centers being driven by artificial intelligence, and that's just in the United States, [indiscernible] again, we do have a global presence here, really drives some of the optimism that we had with the forecast for double-digit growth."<br><br>"We are raising our 2024 outlook to reflect the strong momentum with which we exited 2023. The accelerating pace of investment in workflow automation and interest in Gen AI positions us well on our journey to becoming the defining enterprise software company of the 21st century."<br><br>"U.S. growth of 12.7% was well ahead of our expectations with elective procedure volumes recovering and procedure cancellation rates returning to pre-pandemic levels. "<br><br>"First, is the clearly growing industry-wide box office up in North America by some 29% quarter-to-quarter versus last year, growing to more than $.7 billion in the quarter. " |

| Category | Example Texts from Conference Call Transcripts |
|---|---|
| *Very pessimistic (Score = 1)* | “As economic uncertainty persists, the strength of our model is enabling us to deliver value for our customers, continue to invest in our associates and deliver consistent shareholder return. Value remains top of mind for many of our customers as they are balancing several factors that are impacting their food at home spending. The effect of sustained inflation, reduced government benefits including SNAP, and higher interest rates have pressured customer spending, especially for those on a tight budget."<br><br>“Recent results from across our industry and many of our largest customers has clearly demonstrated that macroeconomic headwinds continue largely unabated. This environment has challenged financial performance of IFF and our peers in the first half of 2023 and has resulted in a more cautious outlook for the remainder of the year."<br><br>“Orders have softened in the residential home furnishings market. I think a lot of that is tied to macroeconomic uncertainty, Steven, I think, also with home sales slowing down. I think we’ll continue to see a softening in demand."<br><br>“Nonetheless, softness and natural gas pricing in the U.S. has had a dampening effect on current rig activity and is contributing to an increased level of contractual churn in the market, not only in terms of number of rigs but also the increased idle time between contracts."<br><br>“Overall market declines in alternative protein consumption including demand decline for plant-based products, persistent supply chain challenges given the macro backdrop since the pandemic and more aggressive inventory management by customers have collectively contributed to pressure Functional Ingredients."<br><br>“Across the globe, different regions are in varying phases of return to office compounded by varying economic conditions and a general slowdown in the housing market. Turning to Global Retail, as I mentioned earlier, we’re seeing a slowdown in the housing market, particularly in luxury." |

**Figure IA.1.** Additional Macroeconomic Variables and AI Scores

Each AI score is constructed with the two-pass procedure described in Section 3.1, where the second pass asks, "Over the next quarter, how does the firm anticipate a change in ..." and returns an integer from 1 to 7 with 4 as the neutral midpoint. The firm-quarter scores are collected each quarter and aggregated into an equally weighted nationwide score. The change in real outcomes are percent change in ... in quarter $t+1$ relative to quarter $t-3$. The change in predicted outcomes are percent change in ... in quarter $t+1$ relative to quarter $t-3$.

**(a)** Aggregate production and AI Production Score

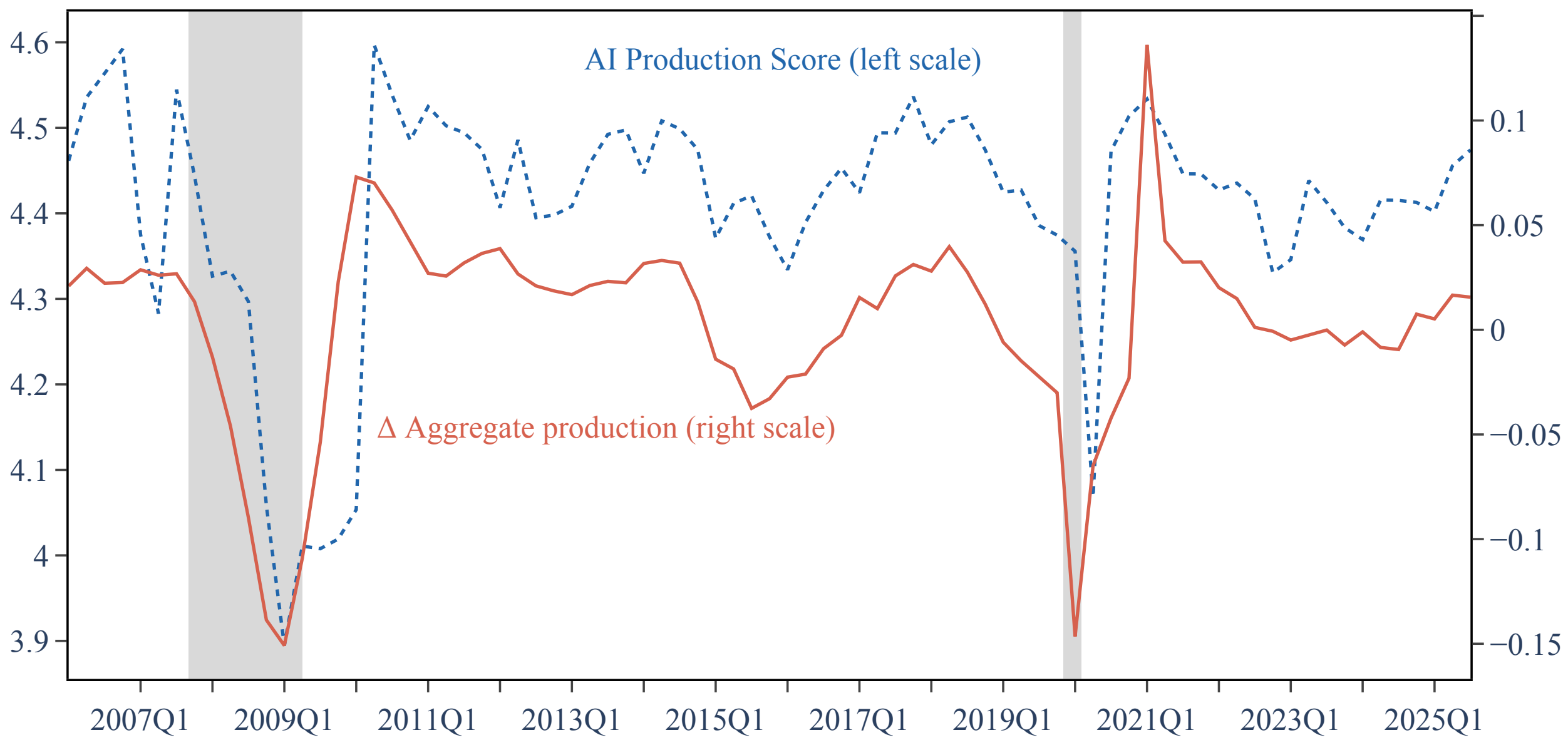

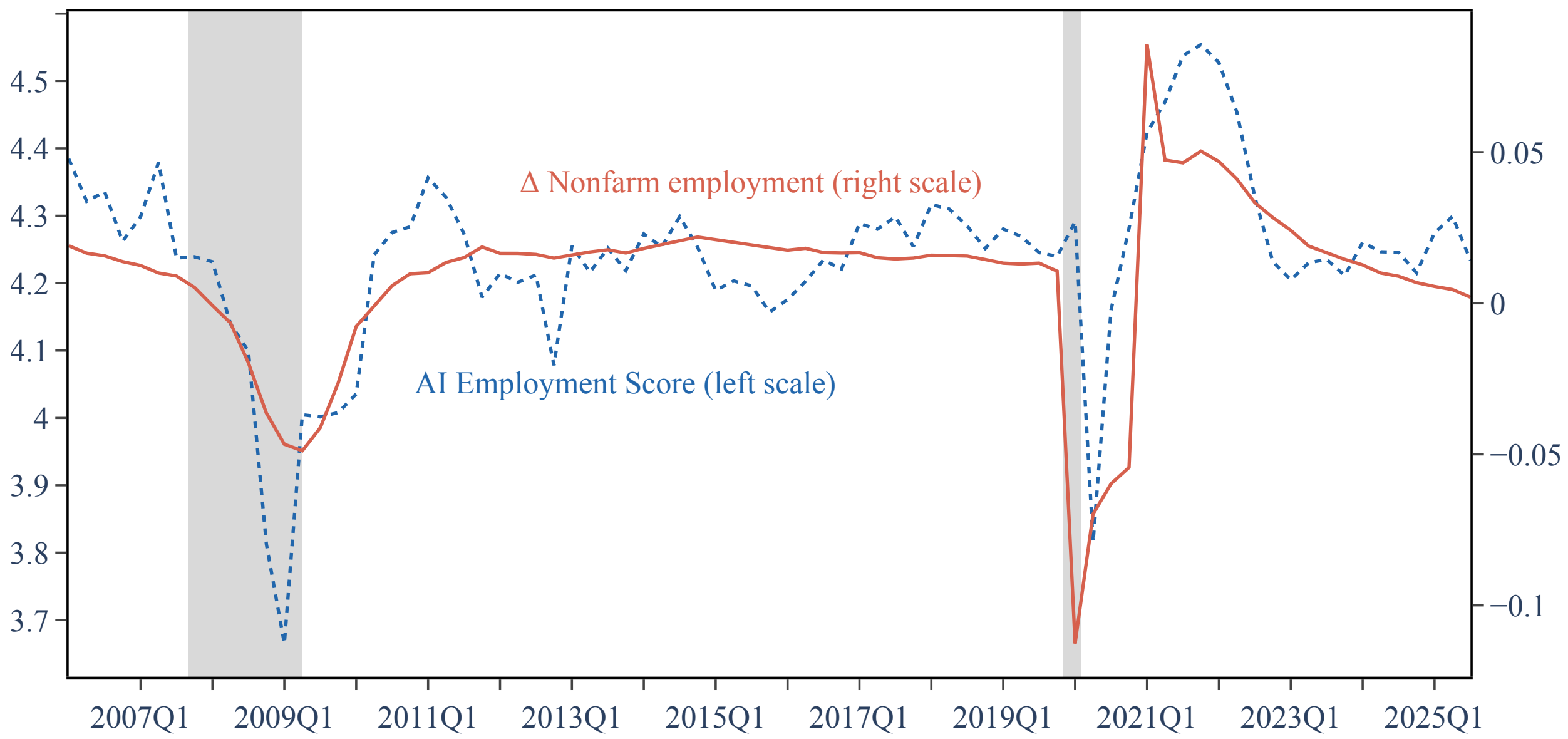

**(b)** Nonfarm payroll employment and AI Employment Score

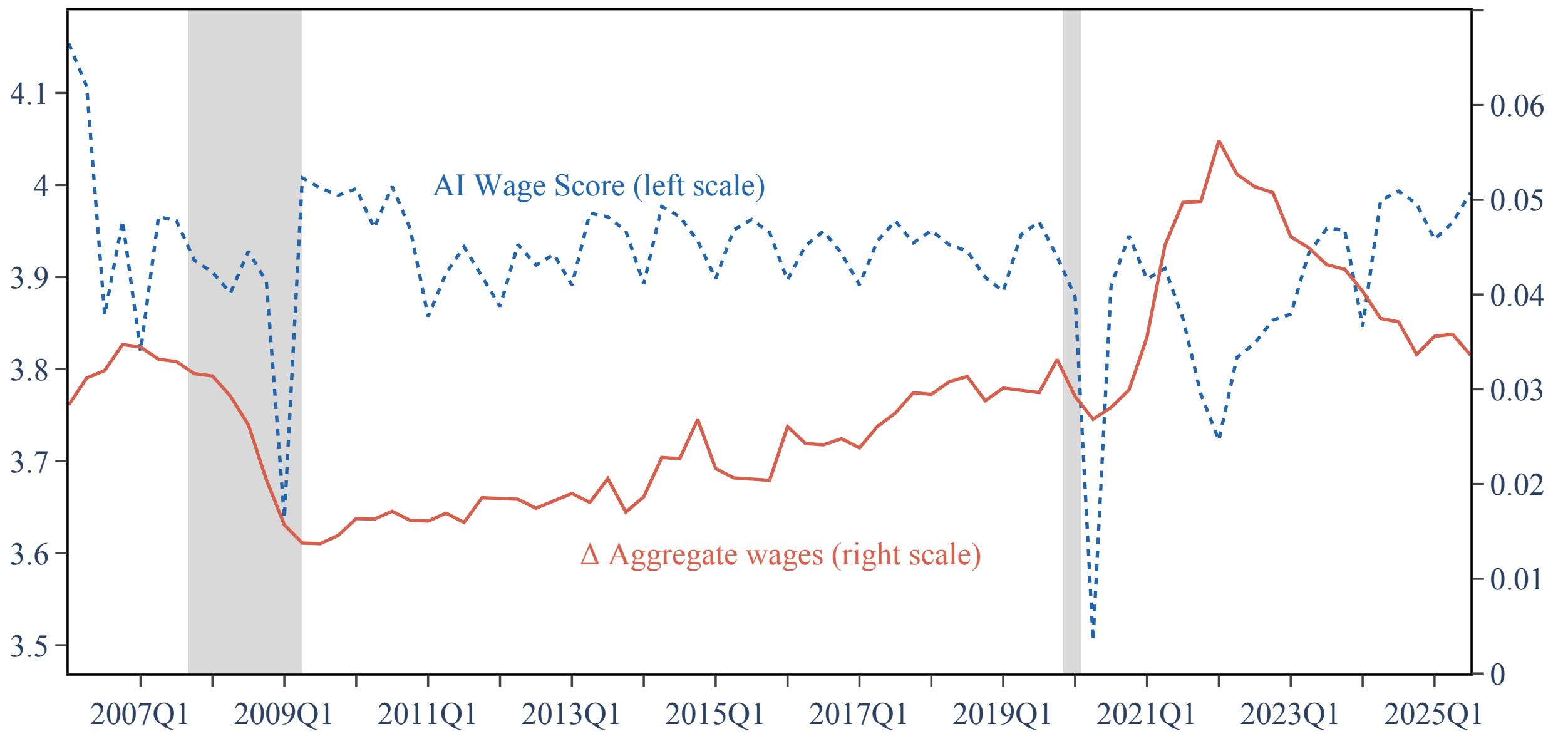

**(c)** Aggregate Wage and AI Wage Score

**Figure IA.2.** AI Production, Employment, and Wage Score across Industry Sectors

**(a)** AI Production Score

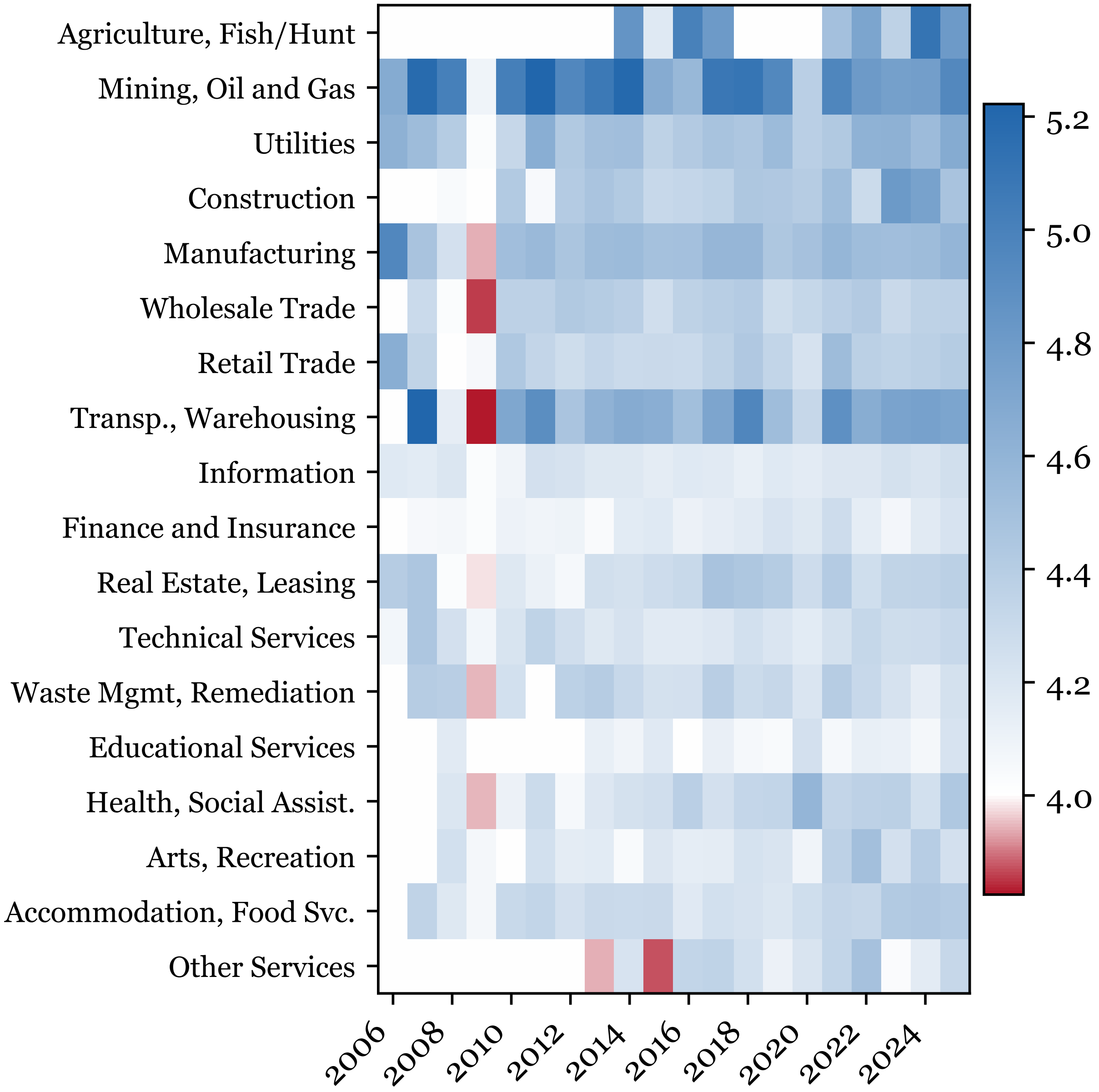

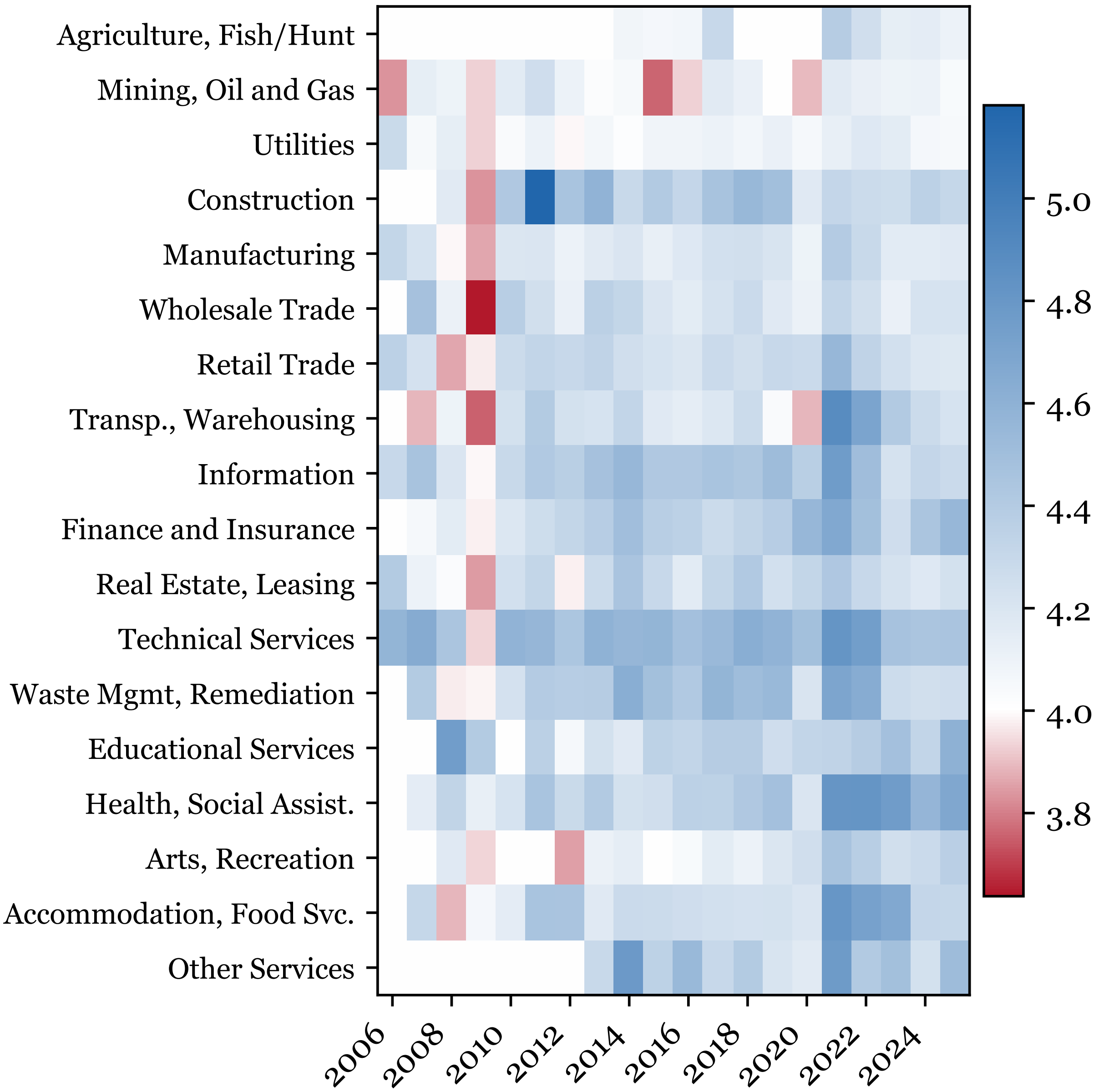

**(b)** AI Employment Score

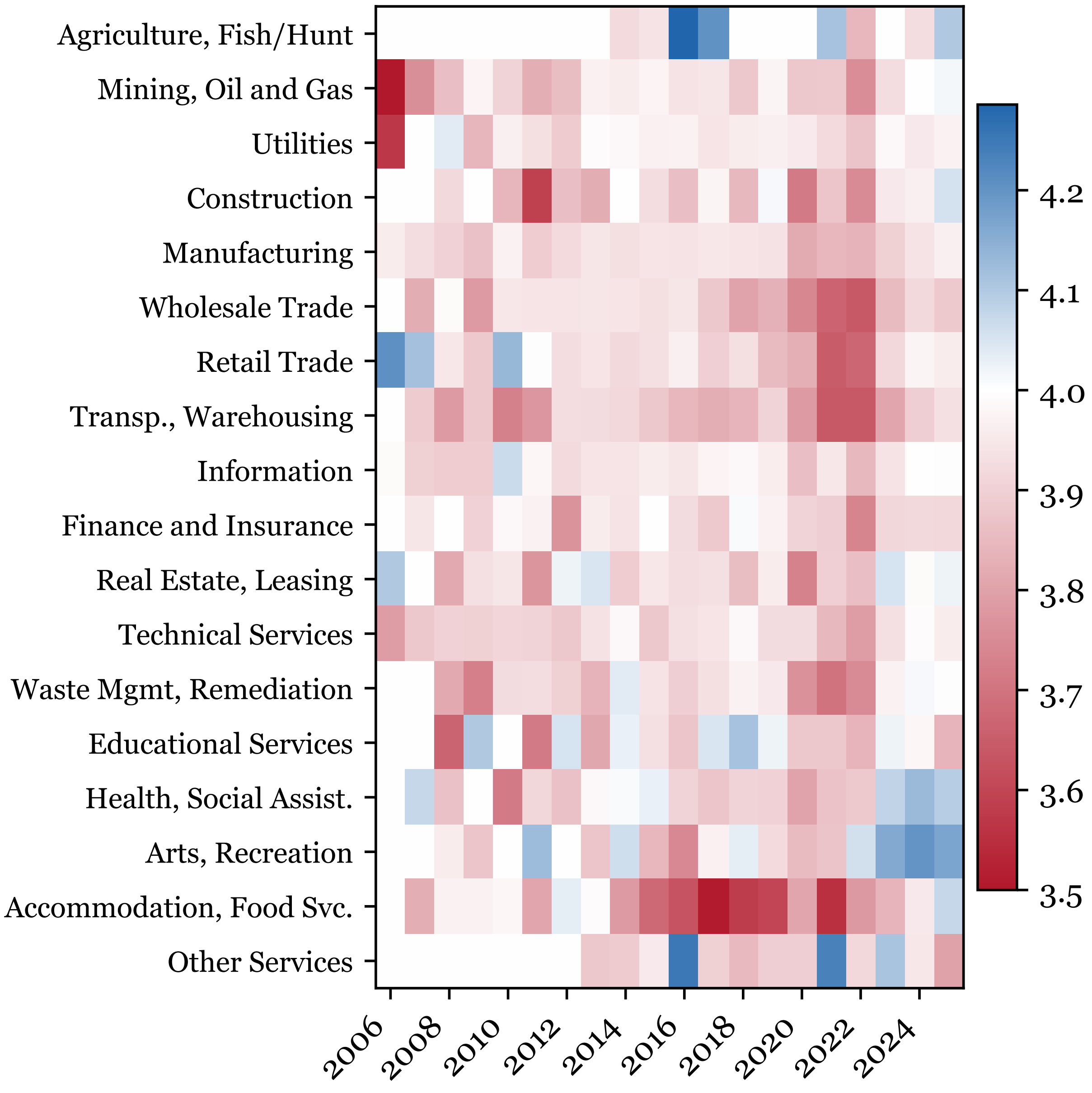

**(c)** AI Wage Score

**Figure IA.3.** Macroeconomic implications of changes in *AI Economy Score*

The figure presents the impulse responses to a one-standard-deviation orthogonal shock to the *AI Economy Score*. The responses of consumption, investment, output growth, output forecast, and excess market return have been accumulated. Shaded areas represent 95-percent confidence intervals, derived from 2,000 bootstrap replications. In each graph, the X-axis shows the quarters after the shock and Y-axis is in percentage points.

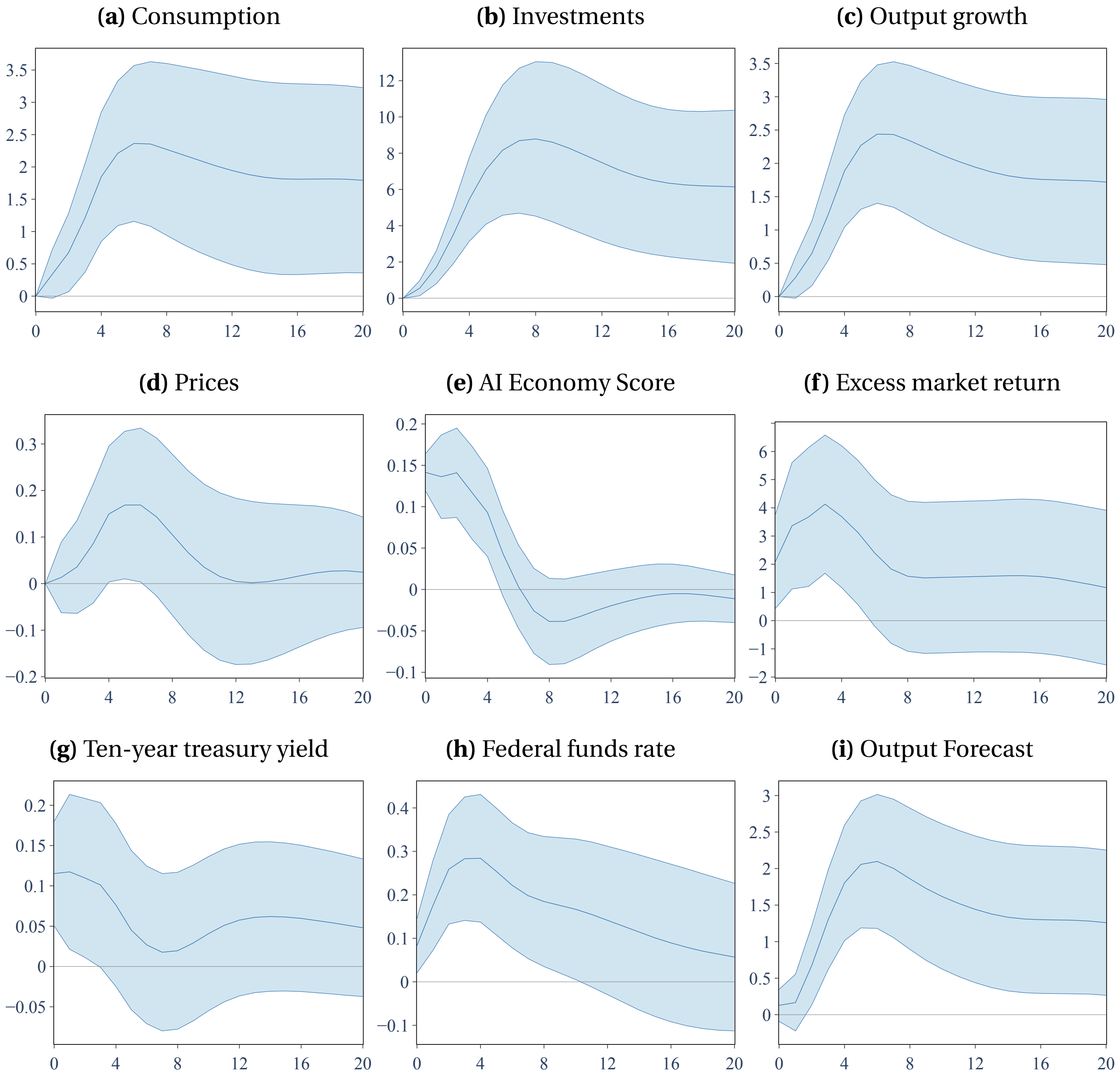

**Table IA.1.** AI predictions on other economic indicators: long horizons at the industry level

This table reports the coefficients from regressing the the growth of economic predictors for long horizons on AI-predicted scores at the industry level. Panel A displays the results on employment. Panel B displays the results on wages. The dependent variables are defined as the logarithm of the next period $t+1$ over the previous period $t-1$. Panel B reports the results for longer horizons. Four lags of the dependent variables are controlled for in all specifications. Variable definitions are in Appendix A.

Panel A: Industry Employment

| | (1) | (2) | (3) |
|---|---|---|---|
| | *Ind. Employment* | | |
| | 2 years | 3 years | 4 years |
| *Term Spread* | 1.207 | -0.0996 | 2.511** |
| | (1.36) | (-0.10) | (2.21) |
| *Real FFR* | -1.184** | -2.174*** | -1.129* |
| | (-2.31) | (-4.33) | (-1.89) |
| *GZ Spread* | -3.162 | 0.409 | -2.211 |
| | (-1.38) | (0.16) | (-0.76) |
| *AI Economy Score* | 0.00847 | 0.0523 | 0.0110 |
| | (0.18) | (1.05) | (0.19) |
| *AI Industry Score* | 0.0237* | 0.0291** | 0.0299** |
| | (1.71) | (1.98) | (2.27) |
| Industry FE | Yes | Yes | Yes |
| R-squared | 0.324 | 0.351 | 0.385 |
| Observations | 297 | 279 | 261 |

*(continued)*

Panel B: Industry Wages

| | (1) | (2) | (3) |
|---|---|---|---|
| | | *Ind. Wages* | |
| | 2 years | 3 years | 4 years |
| *Term Spread* | -2.007* | -4.485*** | -2.496* |
| | (-1.96) | (-3.79) | (-1.75) |
| *Real FFR* | -2.569*** | -4.122*** | -3.418*** |
| | (-4.11) | (-7.12) | (-4.76) |
| *GZ Spread* | -3.947* | 1.994 | -1.325 |
| | (-1.68) | (0.69) | (-0.37) |
| *AI Economy Score* | 0.00597 | 0.101* | 0.0518 |
| | (0.12) | (1.90) | (0.73) |
| *AI Industry Score* | 0.0348** | 0.0343** | 0.0384** |
| | (2.37) | (2.08) | (2.30) |
| Industry FE | Yes | Yes | Yes |
| R-squared | 0.386 | 0.446 | 0.413 |
| Observations | 297 | 279 | 261 |

**Table IA.2.** AI Predictions for Long Horizons: Firm Level

This table reports the coefficients from regressing the growth of the economic predictors for long horizons on AI-predicted scores at the firm level. Panel A presents the results on value-add; Panel B presents results on firm sales; Panel C shows the results on firm employment; Panel D shows the results on firm wages. The dependent variables are defined as the logarithm of the next nth (up to four years) period $t+n$ over the last period $t-1$. Four (Two) lags of dependent variables are controlled for in quarterly (annually) specifications. Variable definitions are in Appendix A.

Panel A: Firm Value-Add

| | (1) | (2) | (3) | (4) | (5) | (6) |
|---|---|---|---|---|---|---|
| | | | *Value-Add* | | | |
| | 2 quarters | 3 quarters | 4 quarters | 8 quarters | 12 quarters | 16 quarters |
| *Term Spread* | 0.928 | -0.561 | -1.269 | -6.746*** | -7.380*** | -6.374*** |
| | (0.74) | (-0.44) | (-0.82) | (-6.05) | (-6.32) | (-4.01) |
| *Real FFR* | -0.409 | -1.314 | -1.952* | -5.670*** | -3.500*** | -1.546 |
| | (-0.56) | (-1.31) | (-1.92) | (-6.32) | (-4.22) | (-1.25) |
| *Book Leverage* | 0.0494** | 0.0604*** | 0.0617** | 0.179*** | 0.198*** | 0.204*** |
| | (2.48) | (2.74) | (2.29) | (4.36) | (3.77) | (3.51) |
| *Size* | -0.0680*** | -0.0917*** | -0.118*** | -0.244*** | -0.333*** | -0.399*** |
| | (-11.46) | (-12.08) | (-11.93) | (-15.59) | (-18.19) | (-17.71) |
| *Tangibility* | -0.219*** | -0.242*** | -0.366*** | -0.199** | -0.284** | -0.354*** |
| | (-3.46) | (-3.76) | (-4.75) | (-2.36) | (-2.62) | (-2.93) |
| *GZ Spread* | -4.111*** | -2.917** | -2.808* | -2.393 | 0.915 | 0.653 |
| | (-3.01) | (-2.49) | (-1.97) | (-1.41) | (0.66) | (0.30) |
| *AI Economy Score* | 0.0327 | 0.0499* | 0.0496 | -0.00780 | 0.0682** | 0.118** |
| | (1.29) | (1.86) | (1.39) | (-0.17) | (2.04) | (2.49) |
| *AI Industry Score* | 0.0763*** | 0.0922*** | 0.110*** | 0.0666*** | 0.0366* | 0.0683*** |
| | (6.14) | (6.72) | (6.79) | (4.32) | (1.95) | (3.67) |
| *AI Firm Score* | 0.0286*** | 0.0299*** | 0.0305*** | 0.0257*** | 0.0188*** | 0.0182*** |
| | (14.59) | (14.04) | (12.96) | (10.64) | (6.19) | (6.80) |
| Firm FE | Yes | Yes | Yes | Yes | Yes | Yes |
| R-squared | 0.357 | 0.244 | 0.343 | 0.435 | 0.512 | 0.569 |
| Observations | 87,496 | 85,368 | 83,004 | 72,077 | 62,291 | 52,960 |

*(continued)*

Panel B: Firm Sales

| | (1) | (2) | (3) | (4) | (5) | (6) |
|---|---|---|---|---|---|---|
| | | | | *Firm Sales* | | |
| | 2 quarters | 3 quarters | 4 quarters | 8 quarters | 12 quarters | 16 quarters |
| *Term Spread* | 0.847 | -0.342 | -1.201 | -7.609*** | -8.557*** | -6.775*** |
| | (0.72) | (-0.28) | (-0.80) | (-6.06) | (-6.73) | (-4.54) |
| *Real FFR* | -0.650 | -1.314 | -2.081** | -6.353*** | -3.899*** | -0.451 |
| | (-0.96) | (-1.31) | (-2.10) | (-7.02) | (-4.49) | (-0.38) |
| *Book Leverage* | 0.000287 | -0.0102 | -0.0129 | 0.0628 | 0.116** | 0.124** |
| | (0.02) | (-0.53) | (-0.50) | (1.60) | (2.48) | (2.21) |
| *Size* | -0.0491*** | -0.0695*** | -0.0937*** | -0.207*** | -0.295*** | -0.354*** |
| | (-9.20) | (-10.94) | (-10.58) | (-14.02) | (-16.05) | (-15.96) |
| *Tangibility* | -0.266*** | -0.280*** | -0.372*** | -0.329*** | -0.378*** | -0.496*** |
| | (-4.85) | (-4.21) | (-5.24) | (-3.66) | (-3.23) | (-3.78) |
| *GZ Spread* | -4.538*** | -3.444*** | -3.094** | -1.916 | 1.735 | 1.421 |
| | (-3.74) | (-2.98) | (-2.60) | (-1.28) | (1.38) | (0.71) |
| *AI Economy Score* | 0.0111 | 0.0346 | 0.0437 | 0.00131 | 0.0655** | 0.127*** |
| | (0.54) | (1.33) | (1.56) | (0.03) | (2.28) | (2.78) |
| *AI Industry Score* | 0.0268* | 0.0292* | 0.0391** | 0.0264* | 0.0167 | 0.0621*** |
| | (1.78) | (1.86) | (2.02) | (1.74) | (1.02) | (3.25) |
| *AI Firm Score* | 0.0190*** | 0.0206*** | 0.0228*** | 0.0220*** | 0.0187*** | 0.0181*** |
| | (14.03) | (14.23) | (14.56) | (12.17) | (7.51) | (7.49) |
| Firm FE | Yes | Yes | Yes | Yes | Yes | Yes |
| R-squared | 0.321 | 0.251 | 0.320 | 0.441 | 0.537 | 0.600 |
| Observations | 87,992 | 86,018 | 83,744 | 72,889 | 63,016 | 53,574 |

*(continued)*

Panel C: Firm Employment

| | (1) | (2) | (3) |
|---|---|---|---|
| | | *Firm Employment* | |
| | 2 years | 3 years | 4 years |
| *Term Spread* | -3.019** | -4.044*** | -2.171 |
| | (-2.50) | (-3.33) | (-1.32) |
| *Real FFR* | -3.285*** | -2.179*** | -0.867 |
| | (-4.62) | (-3.95) | (-0.94) |
| *Book Leverage* | -0.264*** | -0.220*** | -0.196*** |
| | (-4.59) | (-3.30) | (-2.98) |
| *Size* | -0.134*** | -0.226*** | -0.298*** |
| | (-6.18) | (-7.54) | (-7.82) |
| *Tangibility* | -0.640*** | -0.683*** | -0.674*** |
| | (-7.08) | (-6.52) | (-5.28) |
| *GZ Spread* | -0.669 | 1.720 | -2.092 |
| | (-0.39) | (1.17) | (-0.95) |
| *AI Economy Score* | 0.0256 | 0.0714*** | 0.0185 |
| | (0.69) | (3.78) | (0.48) |
| *AI Industry Score* | 0.0450*** | 0.0242 | 0.0340* |
| | (3.35) | (1.51) | (1.76) |
| *AI Firm Score* | 0.0403*** | 0.0341*** | 0.0292*** |
| | (8.00) | (6.26) | (5.08) |
| Firm FE | Yes | Yes | Yes |
| R-squared | 0.433 | 0.522 | 0.594 |
| Observations | 26,864 | 23,518 | 20,248 |

*(continued)*

Panel D: Firm Wages

| | (1) | (2) | (3) |
|---|---|---|---|
| | | *Firm Wages* | |
| | 2 years | 3 years | 4 years |
| *Term Spread* | -0.929 | -2.150 | -3.127 |
| | (-0.51) | (-1.17) | (-0.89) |
| *Real FFR* | -0.868 | -0.639 | 0.588 |
| | (-0.55) | (-0.35) | (0.25) |
| *Book Leverage* | -0.152 | -0.105 | -0.113 |
| | (-1.42) | (-0.85) | (-0.87) |
| *Size* | -0.0530 | -0.116*** | -0.188*** |
| | (-1.58) | (-3.14) | (-4.14) |
| *Tangibility* | -0.500 | -0.363 | -0.430 |
| | (-1.71) | (-1.30) | (-1.29) |
| *GZ Spread* | -2.161 | -0.396 | 1.547 |
| | (-0.54) | (-0.09) | (0.22) |
| *AI Economy Score* | -0.0470 | -0.0158 | 0.0573 |
| | (-0.46) | (-0.18) | (0.37) |
| *AI Industry Score* | 0.0772** | 0.0663* | 0.0499 |
| | (2.85) | (1.87) | (1.17) |
| *AI Firm Score* | 0.0361*** | 0.0135 | 0.0187 |
| | (3.62) | (1.05) | (1.14) |
| Firm FE | Yes | Yes | Yes |
| R-squared | 0.571 | 0.581 | 0.629 |
| Observations | 2,638 | 2,173 | 1,631 |

## Internet Appendix: Prompts and Pipeline Details

This appendix reports the prompts used to construct the AI scores and the mechanical steps applied between the two passes. Section 8 covers the *AI Economy Score*. Section 8 covers the thirteen additional aspects that enter the *AI Weighted Score* of Section 6.3. In both pipelines the model is OpenAI's GPT-5 mini, requests are submitted through the batch endpoint, and transcripts are divided into segments of approximately 16,000 characters, with the split placed at the nearest sentence boundary within the preceding 500 characters so that no sentence is divided across segments.

### A. The *AI Economy Score*

**First pass: extraction.** One request per segment. Each segment is passed as the user message under the following system prompt.

> *The following is a section of a company's earnings call transcript. You are a finance expert. Pay equal attention to both prepared remarks and the Q&A section. Extract sentences where management expresses a VIEW or EXPECTATION about the US economy, not merely about the company's own performance. INCLUDE sentences that explicitly discuss: US macroeconomic conditions or outlook, US consumer demand, spending, or confidence, US employment or hiring trends, US interest rates or monetary policy, US inflation, US recession risk, tariffs and their expected broad economic impact, housing market conditions as a US economic indicator, or guidance that directly cites US economic conditions as a key driver. EXCLUDE the following even if they seem related: backward-looking company financial results (revenue, volume, margins, EPS) reported without an economic forecast, company-specific operational updates (projects, capex, asset integration, cost savings), industry-specific metrics that do not explicitly connect to a broader US economic view, fragmented conversational filler without informational content (e.g. 'You know, it is what it is'), operator or moderator lines, and forward-looking disclaimer boilerplate. A sentence qualifies if a reasonable economist would cite it as evidence of management's view on the US economy. Return only the extracted sentences as a numbered list. Do not summarize or interpret.*

**Intermediate filtering.** The extraction output is cleaned before scoring. When a segment contains no qualifying sentences the model returns prose to that effect rather than an empty response, so we drop any line matching that pattern, including variants such as "no qualifying sentences," "no sentences found," "none of the sentences," and "there are no sentences," together with standalone non-answers such as "none" and "n/a." We strip the list numbering, concatenate the surviving sentences across all segments of a call in the order the segments appear, and renumber them from one. A conference call for which no sentence survives is dropped and does not enter any quarter's average.

**Second pass: scoring.** One request per conference call. The surviving sentences are passed as a numbered list in the user message. The fiscal quarter, derived from the call identifier, and the call date are substituted into the system prompt below.

> *You are a finance expert. The following numbered statements were extracted from a company's earnings call transcript for [fiscal quarter]. Based ONLY on these statements and using only your knowledge as of [call date], assess management's overall tone about the US economy. Do not use knowledge of events after [call date].*
>
> *Question: Looking ahead to the quarter immediately following [fiscal quarter], how optimistic or pessimistic is management about US economic conditions?*

*Guidance:*

*- You may INFER management's view of US economic conditions from commentary about US demand, consumers, housing, labor, input costs, interest rates, credit, industry trends, or their US business, even if the word 'economy' is not used explicitly.*

*- Prefer producing a score whenever the statements support a reasonable inference; reserve null only for cases where there is truly nothing US-related to reason from.*

*- Use 4 (Neutral) when signals are genuinely mixed or roughly offsetting; use 3 or 5 only when the balance clearly leans one way.*

*Rate on this scale:*

*1 = Very pessimistic (expects severe deterioration: recession, crisis, major contraction)*

*2 = Pessimistic (expects meaningful slowdown or worsening conditions)*

*3 = Slightly pessimistic (expects mild headwinds or modest softening)*

*4 = Neutral (no clear directional view, or mixed signals)*

*5 = Slightly optimistic (expects mild tailwinds or modest improvement)*

*6 = Optimistic (expects meaningful strengthening or improving conditions)*

*7 = Very optimistic (expects strong expansion, boom conditions)*

*If the statements contain nothing at all that can be connected to US conditions (e.g., they are purely about non-US operations or entirely non-economic topics), respond with: {"score": null, "explanation": "No US economic conditions mentioned."} Otherwise, respond in JSON format: {"score": <integer 1-7>, "explanation": "<one sentence>"}*

## B. The Thirteen Additional Aspects

The thirteen remaining aspects are extracted and scored jointly rather than one at a time. The topic keys, in the order the model is asked to use, are `global_economy`, `firm_prospects`, `industry_prospects`, `earnings`, `revenue`, `investments`, `cost_of_capital`, `wages_salaries`, `num_employees`, `production_quantity`, `product_demand`, `output_prices`, and `input_prices`.

**First pass: extraction.** One request per segment covering all thirteen topics at once. Issuing a single request rather than one request per topic prevents the same sentence from being extracted under several headings, which cannot be avoided when separate requests are unable to see one another. The system prompt instructs the model to extract sentences in which management expresses a view or expectation about each topic, and imposes five rules: assign each sentence to at most one topic, choosing the more specific one where a sentence could fit two; require an explicit reference to the topic's subject matter, so that generic expressions of confidence do not qualify on their own; exclude operator lines, analyst questions, and disclaimer boilerplate; return `NONE` rather than stretch a sentence to fit; and read the entire segment before assigning, giving later topics the same attention as earlier ones. The model returns one section per topic, headed by `## <topic_key>` on its own line and followed by a numbered list of verbatim sentences, or `NONE`.

The prompt defines each topic by what it includes and what it excludes, with the exclusions drawn to keep neighboring topics apart. To illustrate, `revenue` covers forward-looking statements referencing revenue, sales, top-line growth, comparable sales, or bookings, and excludes statements referencing margins or earnings, which belong to `earnings`, and statements about customer demand without a sales framing, which belong to `product_demand`. `firm_prospects` is a strict residual, reserved for explicit aggregate statements about the firm's overall outlook that fit none of the more specific topics. `industry_prospects` requires the grammatical subject to be the industry, sector, or competitors collectively, so that sentences about the firm's own segments are excluded. `output_prices` covers the prices the firm charges and `input_prices` the prices it pays. `investments` covers capital outlays while `cost_of_capital` covers hurdle rates and return-on-capital metrics. `wages_salaries` requires an explicit labor-cost concept, and `num_employees` requires headcount rather than store or plant counts.

**Second pass: scoring.** The extracted sentences are pooled across all segments of a call and grouped by topic, and one request per conference call scores all thirteen. Topics with no extracted sentences are presented as `NONE`. Calls for which every topic is empty are skipped. The system prompt substitutes the fiscal quarter and call date as above, imposes the same knowledge cutoff, and applies the same seven-point scale to every topic, with the additional instruction to score each topic independently on the basis of its own sentences. The model responds with a single JSON object containing one key per topic, each mapping to a score and a one-sentence explanation, and returns a null score for topics with no sentences or no basis for a directional view.

## Internet Appendix: Anonymization Prompts

Your role is to ANONYMIZE all text that is provided by the user. The text you need to anonymize will be about firm NAME, its ticker, TICKER, and its industry, INDUSTRY. Any instance of the firm, their products, the industry or industry related areas they are involved in should be anonymized. Absolutely no mention of key words related to that industry can be left in the text. After you have anonymized a text, NOBODY, not even an expert financial analyst should be able to use the text to identify the company or the industry the company operates in. For example, if the text is: The country's largest phone producer Apple had great phone-related earnings but Google did not in 2022, likely because of Apple's slogan Think Different, then you should ANONYMIZE it to: The country's largest product_type_1 producer Company_1 had great product_type_1 related earnings but Company_2 did not in time_1 likely because of Company_1's slogan slogan_1. You should also ANONYMIZE any other information which one could use to identify the company or to make an educated guess about its identity. Stock tickers are identifiers, and are usually four or fewer capitalized letters (use TIKR as a stand-in for an arbitrary ticker), and may be referenced in the text in the following formats; SYMBOL:TIKR, $TIKR, >TIKR, $ TIKR, SYMBOL TIKR, SYMBOL: TIKR, > TIKR. Make sure you ANONYMIZE TIKR to ticker_x. Also ANONYMIZE any other identifiers related to companies including the names of individuals, locations, industries, sectors, product names and types and generic product lines, services, times, years, dates and all numbers and percentages in the text, including units. These should be replaced with name_x, location_x, industry_x, sector_x, product_x, product_type_x, product_line_x, service_x, time_x, year_x, date_x and number a, b, c, respectively. Also replace any website or internet links with link_x. You should never just delete an identifier; instead, always replace it with an anonymized analog. After you read and ANONYMIZE the text, you should output the anonymized text and nothing else.

You will receive a body of text which has been anonymized. You are omniscient. Use all available clues in the text to identify which company the text is about, its industry, and the year it was written in. Identify the company, its industry, and the year. Make your best guess if you are unsure. Your options for industry guesses are the following: Aerospace, Agriculture, Automobiles and Trucks, Banking, Beer and Liquor, Construction Materials, Printing and Publishing, Shipping Containers, Business Services, Chemicals, Computers, Apparel, Construction, Coal, Pharmaceutical Products, Electrical Equipment, Fabricated Products, Financial Services, Food Products, Recreation, Precious Metals, Defense, Computer Hardware, Healthcare, Consumer Goods, Insurance, Measuring and Control Equipment, Machinery, Restaurants, Hotels, and Motels, Medical Equipment, Non-Metallic and Industrial Mining, Petroleum and Natural Gas, Paper, Personal Services, Real Estate, Retail, Rubber and Plastic Products, Shipbuilding, Railroad Equipment, Tobacco Products, Candy and Soda, Software, Steel Works, Communication, Entertainment, Transportation, Textiles, Utilities, Wholesale. Provide your estimates EXACTLY in the following format: Ticker Estimate: TIKR,Name Estimate: NAME, Industry Estimate: IND, Year Estimate: Y